# Novel Mixed Approximate Deconvolution Large-Eddy Simulation Models for Two-Phase Flows

Mohammad Bagher Molaei[1], Ehsan Amani[1,*], and Morteza Ghorbani[2]

[1]*Department of Mechanical Engineering, Amirkabir University of Technology (Tehran Polytechnic), Iran*
[2]*Faculty of Engineering and Natural Science, Sabanci University, 34956 Tuzla, Istanbul, Turkey*

## Abstract

Approximate Deconvolution (AD) frameworks for Large-Eddy Simulation (LES) offer a promising approach for modeling multiscale interactions in turbulent multiphase flows. However, some conventional formulations exhibit structural inconsistencies, identified in our previous work [Amani E., Molaei M.B., Ghorbani M. 2024 Novel mixed approximate deconvolution subgrid-scale models for large-eddy simulation. Physics of Fluids 36] which can impact the accuracy and robustness of AD-LES. To address this issue, the present study extends a family of structurally consistent AD closures to turbulent two-phase flows. Specifically, we introduce the Alternative Local Dynamic Model with Kinetic energy equation (A-LDMK), an improved non-equilibrium one-equation mixed eddy-viscosity AD framework. These closures are evaluated through a posteriori analyses of a single-phase turbulent channel flow and a turbulent two-phase gas-liquid mixing layer. The results indicate that the consistent frameworks enhance numerical stability and can sustain energy backscatter under less restrictive stability constraints compared to standard dynamic procedures. Specifically, the proposed A-LDMK provides the most accurate prediction of the multiscale primary atomization cascade in terms of droplet size distribution, compared to the other assessed closures.

[*] Corresponding author. Address: Mechanical Engineering Dept., Amirkabir University of Technology (Tehran Polytechnic), 424 Hafez Avenue, Tehran, P.O.Box: 15875-4413, Iran. Tel: +98 21 64543404. Email: eamani@aut.ac.ir

## 1. Introduction

Turbulent two-phase flows are ubiquitous in both academic research and practical applications of fluid mechanics, encompassing phenomena such as spray dynamics, fuel injection systems, oil pipeline transport, and generation of sea aerosols, among others. The intricate interaction between turbulence and interfacial dynamics often manifests through non-linear effects, leading to macroscopic deformations of the interface, including the formation of ligaments, droplet coalescence, and interface rupture. These complex multiphase interactions pose significant challenges for both experimental and analytical approaches, primarily due to the inherent heterogeneity of the multiphase medium (Vincent *et al.* 2018, Saeedipour & Schneiderbauer 2021). While numerical simulations are highly effective for investigating these unsteady multiphase dynamics system (Labourasse *et al.* 2007, Vincent *et al.* 2008, Larocque *et al.* 2010, Vincent *et al.* 2018, Saeedipour & Schneiderbauer 2021), the extreme grid resolution required by Direct Numerical Simulation (DNS) to capture all interacting scales makes it impractical for real-world applications. Consequently, Large-Eddy Simulation (LES) represents a promising and computationally viable solution. However, unlike LES for single-phase flows, much more challenges still remain to address to achieve robust and reliable LES solutions for multiphase flow problems, primarily due to the intricate interactions of SubGrid-Scale (SGS) physics. This underscores the necessity for ongoing research aimed at developing LES closure models tailored to multiphase LES (Labourasse *et al.* 2007, Larocque *et al.* 2010, Ling *et al.* 2015, Aniszewski 2016, Vincent *et al.* 2018, Saeedipour & Schneiderbauer 2021).

In the field of two-phase LES studies, the conventional approach involves simply using classical single-phase LES frameworks for two-phase flows, neglecting additional closures arising from the presence of the two-phase interface (Srinivasan *et al.* 2008, Chesnel *et al.* 2011, Li *et al.* 2015). However, the application of spatial filtering to the governing equations of two-phase flows—including the Navier-Stokes equations and the transport equation for the phase indicator function—gives rise to several extra unclosed SGS terms (Labourasse *et al.* 2007, Vincent *et al.* 2008, Larocque *et al.* 2010, Aniszewski 2016, Vincent *et al.* 2018, Saeedipour & Schneiderbauer 2021). Among these, only the convective SGS term is shared with single-phase LES.

The critical importance of these uniquely two-phase SGS terms was first established through rigorous *a priori* analyses of DNS data. In a pioneering study, Vincent *et al.* (2008) explicitly filtered their DNS data for a phase separation problem and demonstrated that several of these new terms, especially the SGS contributions from inertia, surface tension, and phase advection, were non-negligible and could reach magnitudes comparable to their resolved counterparts. This pivotal finding highlighted the inadequacy of simply extending single-phase closures to multiphase problems. This conclusion was subsequently confirmed and generalized by Larocque *et al.* (2010) in a detailed parametric study of the same benchmark. By varying density, viscosity, and surface tension, they established an importance hierarchy of the SGS terms, identifying the convective SGS term and the interfacial transport SGS term as the most dominant effects requiring accurate closure models.

With the challenge clearly defined, research bifurcated into two main philosophical approaches for SGS modeling: (1) modifying existing functional models to incorporate multiphase physics, and (2) developing structural models to directly reconstruct the SGS terms. The first path was investigated by Saeedipour and Schneiderbauer (2019), who reported that standard algebraic Eddy-Viscosity (EV) models may not fully account for interfacial turbulence generation. They proposed a modified formulation by including a surface-tension-induced TKE production under the equilibrium assumption used in standard SGS EV models. While this functional approach demonstrated improved predictions for local turbulence characteristics and droplet size distributions (Saeedipour & Schneiderbauer 2021), an alternative approach is to utilize structural or mixed models to directly reconstruct the sub-grid interactions.

The second path, focusing on structural models, gained substantial momentum from a series of compelling *a priori* studies. Vincent *et al.* (2018) performed a comparative assessment of various model types against DNS data. Their results were unequivocal: structural models, particularly the Approximate Deconvolution Model (ADM), exhibited excellent correlation coefficients for all relevant SGS terms (convective, surface tension, and interfacial), whereas functional models like Smagorinsky and WALE showed consistently poor correlations. This superiority of structural approaches was repeatedly confirmed across different multiphase regimes. Ketterl and Klein (2018) and Abdelsayed *et al.* (2024) reached similar conclusions in their analyses of Newtonian and non-Newtonian liquid jet atomization, highlighting the robust performance of scale-similarity and gradient-based models. Furthermore, Schneiderbauer and Saeedipour (2018) demonstrated that the ADM framework also excels in turbulent gas-solid flows, accurately reconstructing unresolved drag forces where functional models failed. Concurrent efforts focused on developing specific structural closures for individual terms, such as the ADM-τ model by Aniszewski (2016) for the SGS surface tension.

Despite their demonstrated *a priori* accuracy, a known deficiency of pure structural models is their tendency to be insufficiently dissipative, which often leads to numerical instability in *a posteriori* simulations. This fundamental conflict between the accuracy of structural models and the robustness of functional models has brought about the use of mixed ADM closures, first proposed by Winckelmans *et al.* (1998) for tensor-diffusivity models and later adapted by Habisreutinger *et al.* (2007) for AD-LES. The practical viability of this approach to two-phase flows was explored through *a posteriori* studies. Saeedipour *et al.* (2019) performed one of the first *a posteriori* simulations using a full two-phase ADM closure demonstrating improved predictions of macroscopic quantities but highlighting the critical challenge of maintaining model robustness. More recently, Iberl *et al.* (2025) conducted an *a posteriori* assessment of mixed models for turbulent polydisperse bubbly flows. Their work showed that by dynamically blending a structural model (Liu-type) with a functional model (Smagorinsky), using either a SGS activity sensor or an explicit scale separation, it is possible to achieve the best of both worlds. These mixed approaches retained the stability of the functional component while leveraging the superior accuracy of the structural component, leading to significantly improved predictions of bubble plume dynamics compared to standalone models. However, despite these advancements, achieving a mathematically consistent and physically faithful closure for complex two-phase regimes remains an active research frontier, particularly regarding the incorporation of flow history and sub-filter scale surface tension effects.

Against this background, the primary objective of the present work is to bridge the gap between the numerical robustness of the consistent mixed AD-LES framework and the physical fidelity required for complex multiphase flows. To this end, three key novelties are introduced. First, a recent highly consistent mixed AD-LES framework developed for single-phase LES (Amani *et al.* 2024) is extended to and evaluated for two-phase flows. Second, we propose a novel one-equation closure model integrated within the mixed AD-LES formulation. By solving a transport equation for the SGS turbulent kinetic energy, this approach explicitly incorporates flow history and non-equilibrium effects—features that are intrinsically absent in standard algebraic closures which rely on the restrictive local equilibrium hypothesis. This physical enhancement is critical for accurately predicting transient atomization processes where local production and dissipation of turbulence are rarely in balance. Third, beyond the momentum closure, this study investigates the contribution of the sub-filter surface tension term. Utilizing the structural ADM for reconstruction, we isolate and quantify the impact of capillary forces at the cut-off scale, a dominant physical mechanism often oversimplified or neglected in conventional functional modeling. The applicable scope of the present framework is focused on the subgrid momentum and surface tension closures. The subfilter phase transport term is omitted at this stage, highlighting a specific area for future research.

The remaining sections of the paper are structured as follows. Section 2 presents the mathematical formulation of the two-phase AD-LES framework and details the proposed SGS closure strategies, including the consistent one-equation model. The information on the benchmarks used to evaluate the method and the numerical algorithms are reported in sections 2.3 and 4, respectively. Section 5 presents the results of the study, including *a posteriori* analyses that carefully assess the accuracy and stability of the closures. Finally, in section 6, the main conclusions and most important findings of the study are summarized.

## 2. Mathematical modeling: Elements of two-phase AD-LES

### *2.1. Two-phase AD-LES governing equations*

The Navier–Stokes equations for two-phase incompressible flows can be written using a one-fluid formulation (Tryggvason *et al.* 2011) as follows:

$$\partial_t(\rho u_i) + \partial_j(\rho u_i u_j) = -\partial_i p + \partial_j(2\mu S_{ij}) + \sigma\kappa\partial_i\alpha + \rho g_i, \tag{1}$$

$$\partial_i(u_i) = 0, \tag{2}$$

$$\partial_t(\alpha) + \partial_i(\alpha u_i) = 0, \tag{3}$$

where $\partial_t$ and $\partial_i$ stand for the time derivative $(\partial/\partial t)$ and spatial derivative $(\partial/\partial x_i)$, respectively, and $u_i$ is the (mixture) velocity vector, $p$ the pressure, $\alpha$ the primary (liquid) phase volume fraction, $\rho$ the density, $\mu$ the viscosity, and $\sigma$ the surface tension coefficient. The strain-rate tensor, $S_{ij}$, and interface curvature, $\kappa$, are given by:

$$S_{ij} = \frac{1}{2}(\partial_j u_i + \partial_i u_j). \tag{4}$$

$$\kappa = \partial_j n_j = \partial_j\left[\frac{\partial_j\alpha}{(\partial_k\alpha\partial_k\alpha)^{1/2}}\right] \equiv \kappa(\alpha). \tag{5}$$

Here, $n_i$ is the interface normal vector. The properties without a subscript refer to the mixture properties and are computed by:

$$\rho = \rho_l \alpha + \rho_v (1 - \alpha), \tag{6}$$

$$\mu = \mu_l \alpha + \mu_v (1 - \alpha). \tag{7}$$

When solving equations on an LES grid, a grid filter, $\widetilde{(.)}$, is implicitly applied to Eqs. (1)-(7):

$$\partial_t(\tilde{\rho}\tilde{u}_i) + \partial_t \tau_i^{u\rho} + \partial_j(\tilde{\rho}\tilde{u}_i\tilde{u}_j) + \partial_j \tau_{ij}^{uu} = -\partial_i \tilde{p} + \partial_j(2\tilde{\mu}\tilde{S}_{ij}) + \partial_j \tau_{ij}^{u\mu} + \sigma\tilde{\kappa}\partial_i\tilde{\alpha} + \tau_i^{u\sigma} + \tilde{\rho}g_i, \tag{8}$$

$$\partial_i \tilde{u}_i = 0, \tag{9}$$

$$\partial_t \tilde{\alpha} + \partial_i(\tilde{u}_i\tilde{\alpha}) + \partial_i \tau_i^{\alpha u} = 0, \tag{10}$$

$$\tilde{S}_{ij} = \frac{1}{2}(\partial_j \tilde{u}_i + \partial_i \tilde{u}_j), \tag{11}$$

$$\tilde{\rho} = \rho_l \tilde{\alpha} + \rho_v (1 - \tilde{\alpha}), \tag{12}$$

$$\tilde{\mu} = \mu_l \tilde{\alpha} + \mu_v (1 - \tilde{\alpha}), \tag{13}$$

$$\tilde{\kappa} \equiv \kappa(\tilde{\alpha}), \tag{14}$$

where $\tau_i^{u\rho}$, $\tau_{ij}^{uu}$, $\tau_{ij}^{u\mu}$, $\tau_i^{u\sigma}$, and $\tau_i^{\alpha u}$ are the SGS tensors and vectors defined by:

$$\tau_i^{u\rho} = \widetilde{\rho u_i} - \tilde{\rho}\tilde{u}_i, \tag{15}$$

$$\tau_{ij}^{uu} = \widetilde{\rho u_i u_j} - \tilde{\rho}\tilde{u}_i\tilde{u}_j, \tag{16}$$

$$\tau_{ij}^{u\mu} = \widetilde{2S_{ij}\mu} - 2\tilde{S}_{ij}\tilde{\mu}, \tag{17}$$

$$\tau_i^{u\sigma} = \widetilde{\sigma\kappa\partial_i\alpha} - \sigma\tilde{\kappa}\partial_i\tilde{\alpha}, \tag{18}$$

$$\tau_i^{\alpha u} = \widetilde{u_i\alpha} - \tilde{u}_i\tilde{\alpha}. \tag{19}$$

While spatial filtering yields several SGS terms, Eqs. (15)-(19), they do not contribute equally to the overall flow dynamics. In this study, we focus on the terms widely reported as dominant in multiphase LES: the SGS inertia tensor ($\tau_{ij}^{uu}$), the SGS surface tension ($\tau_i^{u\sigma}$), and the SGS phase transport ($\tau_i^{\alpha u}$) (Larocque et al. 2010, Vincent et al. 2018). The SGS viscous stress tensor ($\tau_{ij}^{u\mu}$) is neglected, since its contribution is assumed to be small compared to the convective term ($\tau_{ij}^{uu}$) at the high Reynolds numbers under consideration. Furthermore, the SGS density-velocity correlation term ($\tau_i^{u\rho}$) is also assumed to be negligible (Saeedipour et al. 2019). It is worth noting that the choice of the classical spatial filtering over density-weighted (Favre) approach and the subsequent neglect of specific SGS terms, for the present AD-LES, were guided by recent *a priori* studies of primary atomization and shear-driven multiphase flows. While Favre filtering is frequently adopted in two-phase flows because it analytically absorbs the SGS density-velocity correlation (Abbas *et al.* 2021, Saeedipour & Schneiderbauer 2021), it complicates the deconvolution of the velocity and volume fraction fields. Furthermore, classical filtering preserves the structure of the phase transport equation, allowing for a direct modeling of the SGS interfacial dynamics (Klein *et al.* 2019). Therefore, classical spatial filtering remains a standard approach for two-phase LES, particularly when structural closures like AD are employed (Ketterl & Klein 2018, Klein *et al*. 2019).

In frameworks relying exclusively on an implicit grid filter, the non-dimensional cut-off wavenumber ($\omega'_C$) is the same as the non-dimensional Nyquist wavenumber ($\omega'_N = \pi$). A critical issue in this approach is the potential for numerical discretization errors to corrupt wavenumbers approaching $\omega'_N$. To mitigate this contamination, the AD-LES technique incorporates an extra explicit filter. The purpose of this filter is to define the cut-off wavenumber such that $\omega'_C \ll \omega'_N$, thereby establishing a distinct separation between the resolved scales and the numerically-dominated high-wavenumber region. This explicit filtering operation, represented here by $\overline{(.)}$ or $G * (.)$ with a filter width of width $\bar{\Delta}$ (usually chosen as $\bar{\Delta} \sim 2\tilde{\Delta} \sim 2\Delta$ where $\Delta$ is the local LES grid size), is applied to the governing Eqs. (8)-(19). In this filtering process, Sub-Filter-Scale (SFS) terms arise similar to the SGS terms. Under the assumption of negligible SFS viscous stress and density-velocity correlation, similar to their SGS counterparts, the final set of equations are written as follows:

$$\partial_t(\bar{\tilde{\rho}}\bar{\tilde{u}}_i) + \partial_j\left(\bar{\tilde{\rho}}\bar{\tilde{u}}_i\bar{\tilde{u}}_j\right) + \partial_j T^{uu}_{ij} = -\partial_i\bar{\tilde{p}} + \partial_j\left(2\bar{\tilde{\mu}}\bar{\tilde{S}}_{ij}\right) + \sigma\bar{\tilde{\kappa}}\partial_i\bar{\tilde{\alpha}} + T^{u\sigma}_i + \bar{\tilde{\rho}}g_i, \tag{20}$$

$$\partial_i\bar{\tilde{u}}_i = 0, \tag{21}$$

$$\partial_t\bar{\tilde{\alpha}} + \partial_i(\bar{\tilde{u}}_i\bar{\tilde{\alpha}}) + \partial_i T^{\alpha u}_i = 0, \tag{22}$$

$$\bar{\tilde{S}}_{ij} = \frac{1}{2}\left(\partial_j\bar{\tilde{u}}_i + \partial_i\bar{\tilde{u}}_j\right), \tag{23}$$

$$\bar{\tilde{\rho}} = \rho_l\bar{\tilde{\alpha}} + \rho_v(1-\bar{\tilde{\alpha}}), \tag{24}$$

$$\bar{\tilde{\mu}} = \mu_l\bar{\tilde{\alpha}} + \mu_v(1-\bar{\tilde{\alpha}}), \tag{25}$$

$$\bar{\tilde{\kappa}} \equiv \kappa(\bar{\tilde{\alpha}}), \tag{26}$$

where $T^{uu}_{ij}$, $T^{u\sigma}_i$, and $T^{\alpha u}_i$ are, respectively, the unclosed SFS inertia tensor, SFS surface tension vector, and SFS phase transport vector, defined by:

$$T^{uu}_{ij} = \overline{\widetilde{\rho u_i u_j}} - \bar{\tilde{\rho}}\bar{\tilde{u}}_i\bar{\tilde{u}}_j = \left(\overline{\tilde{\rho}\tilde{u}_i\tilde{u}_j} - \bar{\tilde{\rho}}\bar{\tilde{u}}_i\bar{\tilde{u}}_j\right) + \left(\overline{\widetilde{\rho u_i u_j} - \tilde{\rho}\tilde{u}_i\tilde{u}_j}\right) \equiv b^{uu}_{ij} + a^{uu}_{ij}, \tag{27}$$

$$T^{u\sigma}_i = \overline{\sigma\widetilde{\kappa\partial_i\alpha}} - \sigma\bar{\tilde{\kappa}}\partial_i\bar{\tilde{\alpha}} = \left(\overline{\sigma\tilde{\kappa}\partial_i\tilde{\alpha}} - \sigma\bar{\tilde{\kappa}}\partial_i\bar{\tilde{\alpha}}\right) + \left(\overline{\sigma\widetilde{\kappa\partial_i\alpha} - \sigma\tilde{\kappa}\partial_i\tilde{\alpha}}\right) \equiv b^{u\sigma}_i + a^{u\sigma}_i, \tag{28}$$

$$T^{\alpha u}_i = \overline{\widetilde{u_i\alpha}} - \bar{\tilde{u}}_i\bar{\tilde{\alpha}} = \left(\overline{\tilde{u}_i\tilde{\alpha}} - \bar{\tilde{u}}_i\bar{\tilde{\alpha}}\right) + \left(\overline{\widetilde{u_i\alpha} - \tilde{u}_i\tilde{\alpha}}\right) \equiv b^{\alpha u}_i + a^{\alpha u}_i, \tag{29}$$

Here, each term is decomposed into a deconvolved SFS, $b$, and a modeled SFS, $a$, component, defined by the terms in the first and second parentheses, respectively.

In the AD-LES framework, Eqs. (20)-(26) are solved directly for the (doubly-)filtered fields, $\bar{\tilde{u}}_i$, $\bar{\tilde{p}}$, and $\bar{\tilde{\alpha}}$. This objective necessitates a closure for the unclosed SFS terms, $T^{uu}_{ij}$, $T^{u\sigma}_i$, and $T^{\alpha u}_i$.

In AD-LES, the deconvolved SFS component, $b$, is expressed as a function of $\tilde{u}_i$ (or $\tilde{\alpha}$), which is approximately recovered through the "soft" deconvolution, i.e., an approximation of the inverse filter. Conversely, the modeled SFS component, $a$, contains the remaining unresolved scales and must be explicitly approximated by an SGS closure, the "hard" deconvolution.

To reconstruct the required field $\tilde{u}_i$ (or $\tilde{\alpha}$), the Van Cittert iterative algorithm (Van Cittert 1931) is employed. This method is predicated on expanding the inverse explicit filter operator, $G^{-1}$. For practical implementation, this series is truncated at the $N$-th term, yielding $Q_N$ as a computable approximation of $G^{-1}$. Following Maulik and San (2018a), the recursive formulation of this algorithm is expressed as:

$$\begin{aligned} &\phi_0^{\star} = \bar{\bar{\phi}}, \\ &\phi_n^{\star} = \phi_{n-1}^{\star} + \left(\bar{\bar{\phi}} - G * \phi_{n-1}^{\star}\right) \ ; n = 1,2, \dots, N, \end{aligned} \tag{30}$$

where $\phi^{\star}$ is the (deconvolved) approximation to $\tilde{\phi}$. Here, $N = 5$ is chosen.

For the soft-deconvolution stage, the deconvolved SFS terms in Eqs. (27)-(29) are modeled according to the generalized Scale-Similarity ADM (SSADM) framework (Stolz & Adams 1999) as:

$$b_{ij}^{uu} \approx \overline{\rho^{\star} u_i^{\star} u_j^{\star}} - \overline{\rho^{\star} u_i^{\star}}\, \overline{u_j^{\star}}, \tag{31}$$

$$b_i^{u\sigma} \approx \overline{\sigma \kappa^{\star} \partial_i \alpha^{\star}} - \sigma \overline{\kappa^{\star}} \partial_i \overline{\alpha^{\star}}, \tag{32}$$

$$b_i^{\alpha u} \approx \overline{u_i^{\star} \alpha^{\star}} - \overline{u_i^{\star}}\,\overline{\alpha^{\star}}, \tag{33}$$

where all $\star$-superscripted variables are computed using the Van Cittert deconvolution method, with the exception of $\rho^{\star}$ and $\kappa^{\star}$, which are determined directly by:

$$\rho^{\star} = \rho_l \alpha^{\star} + \rho_v (1 - \alpha^{\star}), \tag{34}$$

$$\kappa^{\star} \equiv \kappa(\alpha^{\star}). \tag{35}$$

### *2.2. AD-LES SGS Models: The hard deconvolution*

This section details the closure models required for the "hard deconvolution" SFS terms ($a$) introduced in Eqs. (27)-(29). A key methodological assumption is made at this stage: the modeled parts of SFS terms associated with surface tension ($a_i^{u\sigma}$) and phase transport ($a_i^{\alpha u}$) are considered negligible in contrast to their deconvolved counterparts ($b_i^{u\sigma}$ and $b_i^{\alpha u}$). Consequently, the closure problem is simplified to modeling only the modeled SFS momentum inertia tensor, $a_{ij}^{uu}$.

To provide this closure, the S-DEV model (Habisreutinger et al. 2007) can be used, which has been used in all previous mixed AD-LES studies. However, in our previous publication (Amani *et* *al.* 2024), we introduced five mathematical constraints (CI-CV) on SFS stress models used in AD-LES, reflecting the mathematical properties of the exact SFS stresses in limiting conditions that a "consistent" modeling closure is expected to satisfy. The primary consistency criterion, referred to as CI, evaluates the closure's behavior in the theoretical limit of exact deconvolution. In this ideal limit, the deconvolution error diminishes and the deconvolved velocity ($\boldsymbol{u}^{\star}$) approaches the filtered velocity ($\tilde{\boldsymbol{u}}$). However, the exact mathematical definition of the modeled SFS stress ($a_{ij}$), representing flow scales smaller than the grid size, would not approach the identically zero function, since this stress is primarily responsible for maintaining the SGS dissipation of TKE within the whole model. We mathematically proved (Amani *et al.* 2024) that conventional pure AD closures, like the penalty-term regularizations (Stolz *et al.* 2001, Sagaut 2006), or mixed models, such as the Standard Dynamic EV (S-DEV) (Habisreutinger *et al.* 2007), do not respect this constraint. Here, we *a posteriori* revisit this unphysical behavior of S-DEV for single- and two-phase flow benchmarks in the results and discussion section. To resolve the primary consistency issue, we proposed a new family of models, including the Alternative Dynamic Eddy-Viscosity (A-DEV) framework. The A-DEV approach fundamentally differs by modeling the original SGS stress ($\tau_{ij}^{uu}$) and then explicitly filtering the result to obtain $a_{ij}^{uu}$, in contrast to S-DEV which directly models the modeled SFS stress, $a_{ij}^{uu}$.

Given the enhanced consistency, superior stability, and computational efficiency demonstrated in the original study, the A-DEV framework could serve as a more robust and reliable foundation for two-phase flow problems. Furthermore, an important contribution of the present work is the development and evaluation of a novel one-equation model that integrates this consistent A-DEV framework with a transport equation for turbulent kinetic energy (TKE), in addition to the algebraic closures used in our previous work (Amani *et al.* 2024). All aforementioned models are described in the following sub-sections.

#### 2.2.1. *The Standard Dynamic Eddy-Viscosity (S-DEV) closure*

The conventional approach in the literature, which we term Standard DEV (S-DEV), applies the DEV model directly to the modeled SFS stress, $a_{ij}^{uu}$. In the S-DEV framework, the deviatoric component of $a_{ij}^{uu}$ , indicated by the superscript $r$, is modeled as (Habisreutinger *et al.* 2007):

$$-a_{ij}^{uu,r} = C_{\mathrm{D}} m_{ij}(\bar{\tilde{u}}_l, \bar{\bar{\rho}}); \;\; a_{ij}^{uu,r} = a_{ij}^{uu} - \frac{1}{3} a_{kk}^{uu} \delta_{ij}, \tag{36}$$

$$m_{ij}(\bar{\tilde{u}}_l, \bar{\bar{\rho}}) = 2\bar{\bar{\rho}}\bar{\bar{\Delta}}^2 \bar{\tilde{S}} \bar{\tilde{S}}_{ij}^{r}, \tag{37}$$

where $C_{\mathrm{D}}$ is the DEV coefficient. Using the SSADM soft deconvolution, Eq. (31), the model coefficient is calculated by:

$$C_{\mathrm{D}} = \frac{P_{ij}^{r} M_{ji}}{M_{mn} M_{nm}}, \tag{38}$$

$$P_{ij} = \left( \widehat{\bar{\tilde{\rho}} \bar{\tilde{u}}_i \bar{\tilde{u}}_j} - \hat{\bar{\tilde{\rho}}} \hat{\bar{\tilde{u}}}_i \hat{\bar{\tilde{u}}}_j \right) - \left( \widehat{\overline{\rho^\star}\, \overline{u_i^\star}\, \overline{u_j^\star}} - \widehat{\overline{\rho^\star}}\, \widehat{\overline{u_i^\star}}\, \widehat{\overline{u_j^\star}} \right), \tag{39}$$

where $\widehat{(.)}$ is an explicit test filtering of width $\hat{\Delta} \sim 2\bar{\Delta} \sim 4\tilde{\Delta}$ and the same type as the explicit filter $\overline{(.)}$. To complete the modeling, $M_{ij}$ in Eq. (38) is defined based on the chosen DEV closure. Here, this term is expressed using the two algebraic closures, the Standard Dynamic Smagorinsky Model (S-DSM) and Standard Local Dynamic Model with Equilibrium assumption (S-LDME) (Mokhtarpoor & Heinz 2017) as:

$$M_{ij} = \widehat{m_{ij}(\bar{\tilde{u}}_l)} - m_{ij}\left(\hat{\bar{\tilde{u}}}_l\right); \quad (\mathrm{DSM}), \tag{40}$$

$$M_{ij} = -m_{ij}\left(\hat{\bar{\tilde{u}}}_l\right); \quad (\mathrm{LDME}), \tag{41}$$

The aforementioned S-DSM mixed AD-LES approach has been utilized in many previous works, e.g., (Habisreutinger *et al.* 2007, Maulik & San 2018b, Saeedipour *et al.* 2019).

### 2.2.2. *The Alternative Dynamic Eddy-Viscosity (A-DEV) closures*

In contrast to the former approach described in section 2.2.1, a more consistent family of closures, called Alternative DEV (A-DEV), for $a_{ij}^{uu}$ can be derived (Amani *et al.* 2024) which is extended here to the variable density and viscosity condition appropriate for the two-phase formulation as:

$$a_{ij}^{uu,r} = \bar{\tau}_{ij}^{uu,r} = G * \tau_{ij}^{uu,r}, \tag{42}$$

$$-\tau_{ij}^{uu,r} = 2\mu_r \tilde{S}^r{}_{ij} = 2\mu_r S_{ij}^{\star,r}, \tag{43}$$

$$\mu_r = C_{\mathrm{D}} \rho^\star \tilde{\Delta}^{1+\gamma} S^{\star\gamma} q^{\star\frac{1-\gamma}{2}}, \nu_r = C_{\mathrm{D}} \tilde{\Delta}^{1+\gamma} S^{\star\gamma} q^{\star\frac{1-\gamma}{2}}, \tag{44}$$

$$-\tau_{ij}^{uu,r} = C_{\mathrm{D}} m_{ij}(\tilde{u}_l, \tilde{\rho}) = C_{\mathrm{D}} m_{ij}(u_l^\star, \rho^\star), \tag{45}$$

$$m_{ij}(u_l^\star, \rho^\star) = 2\rho^\star \tilde{\Delta}^{1+\gamma} S^{\star\gamma} q^{\star\frac{1-\gamma}{2}} S_{ij}^{\star,r}, \tag{46}$$

where the deconvolved velocity and density field, $u_i^\star$ and $\rho^\star$, directly obtained by the Van Cittert iterative algorithm are used overall this closure formulation. To obtain the dynamic coefficient, $C_{\mathrm{D}}$, starting from Eq. (8) and applying a test box filter, $\breve{(.)}$, as an approximation to the grid filter type and twice its width ($\breve{\Delta} \sim \breve{\tilde{\Delta}} = 2\tilde{\Delta}$):

$$\partial_t\left(\breve{\tilde{\rho}} \breve{\tilde{u}}_i\right) + \partial_j\left(\breve{\tilde{\rho}} \breve{\tilde{u}}_i \breve{\tilde{u}}_j\right) + \partial_j \tau_{ij}^{T,uu} = -\partial_i \breve{\tilde{p}} + \partial_j \left(2\breve{\tilde{\mu}} \breve{\tilde{S}}_{ij}\right) + \sigma \breve{\tilde{\kappa}} \partial_i \breve{\tilde{\alpha}} + \breve{\tilde{\rho}} g_i, \tag{47}$$

$$\tau_{ij}^{T,uu} = \breve{\widetilde{\rho u_i u_j}} - \breve{\tilde{\rho}} \breve{\tilde{u}}_i \breve{\tilde{u}}_j. \tag{48}$$

Defining the Leonard stress (Leonard 1974), $L_{ij} = \tau_{ij}^{T,uu} - \breve{\tau}_{ij}^{uu}$, modeling $\tau_{ij}^{T,uu}$ based on Eq. (45), i.e., $\tau_{ij}^{T,M} = C_{\mathrm{D}} m_{ij}(\breve{\tilde{u}}_l) = C_{\mathrm{D}} m_{ij}(\breve{u_l^\star})$, and using the least-squares error minimization, the dynamic coefficient in Eq. (45) can be computed by:

$$C_{\mathrm{D}} = \frac{L_{ij}^r M_{ji}}{M_{mn} M_{nm}}, \tag{49}$$

$$L_{ij} = \breve{\tilde{\rho} \tilde{u}_i \tilde{u}_j} - \breve{\tilde{\rho}}\, \breve{\tilde{u}}_i \breve{\tilde{u}}_j \approx \rho^\star \breve{u_i^\star u_j^\star} - \breve{\rho^\star} \breve{u_i^\star} \breve{u_j^\star}, \tag{50}$$

To complete the dynamic procedure (Germano *et al.* 1991), the tensor $M_{ij}$ in Eq. (49) must be given based on a selected DEV closure. This framework accommodates both algebraic and TKE-based models by interpreting the variable $q^\star$ and the exponent $\gamma$ in Eq. (44).

For the algebraic closures, A-DSM and A-LDME, the SFS viscosity is assumed to be a direct function of the strain rate, which is achieved by setting $\gamma = 1$ in Eq. (44). This effectively removes the dependency on $q^\star$. The $M_{ij}$ tensor is then formulated by applying the Germano-Lilly identity to the $m_{ij}$ tensor:

$$M_{ij} = \breve{m_{ij}(u_l^\star, \rho^\star)} - m_{ij}(\breve{u_l^\star}, \breve{\rho^\star}); \text{ (A-DSM)}, \tag{51}$$

$$M_{ij} = -m_{ij}(\breve{u_l^\star}, \breve{\rho^\star}); \text{ (A-LDME)}, \tag{52}$$

For the one-equation model, called A-LDMK barrowed from the Local Dynamic Model with Kinetic energy equation (LDMK) proposed by Mokhtarpoor and Heinz (2017) for pure EV-LES, we set $\gamma = 0$. In contrast to LDMK, AD-LES A-LDMK model requires the unresolved turbulent kinetic energy at two distinct scales, namely $q^\star$ (at the model level, used in $m_{ij}$) and $\breve{q^\star}$ (at the test-filter level, used in $M_{ij}$). We identify the deconvolved TKE $q^\star$ with the SGS kinetic energy, $k^\star \equiv k_{SGS}$. For calculation of TKE at the test-filter level, we employ a closure assumption that relates $\breve{q^\star}$ to the trace of the resolved Leonard stress (Eq. (50)):

$$\breve{q^\star} = \max\left(\frac{L_{nn}}{2\breve{\rho^\star}}, 0\right), \tag{53}$$

This physically-based assumption—that the test-level TKE is represented by the resolved stresses at that level—provides a reasonable closure. The $M_{ij}$ tensor for the A-LDMK model, which follows the LDME formulation, is thus defined as:

$$M_{ij} = -m_{ij}(\breve{u_l^\star}, \breve{\rho^\star}, \breve{q^\star}); \text{ (A-LDMK)}, \tag{54}$$

where $m_{ij}$ is computed from Eq. (46) with $\gamma = 0$ and $\widetilde{q^{\star}}$ is computed from the closure in Eq. (53). The model-level TKE, $k^{\star}$, is obtained by solving a transport equation (Lilly 1967, Deardorff 1973) (derived in Appendix A):

$$\partial_t(\rho^{\star}k^{\star}) + \partial_j\left(\rho^{\star}u_j^{\star}k^{\star}\right) = \partial_j\left[\left(\mu + \mu_r\right)\partial_j k^{\star}\right] + \mu_r S^{\star 2} - \frac{\rho^{\star}k^{\star\frac{3}{2}}}{\tilde{\Delta}}. \tag{55}$$

*2.3. AD-LES: statistics computation*

Flow statistics are computed via ensemble (Reynolds) averaging, denoted by $\langle . \rangle$. Key quantities include the mean velocity $\langle u_i \rangle$ and the Reynolds stress, $\langle u_i' u_j' \rangle$, where $u_i' = u_i - \langle u_i \rangle$ is the fluctuating velocity component. Based on the common assumption that $\langle q \rangle \approx \langle \bar{\tilde{q}} \rangle$ (Pope 2000), the mean velocity and mean strain rate are directly approximated by:

$$\langle u_i \rangle \approx \langle \bar{\tilde{u}}_i \rangle; \ \langle S_{ij} \rangle \approx \langle \bar{\tilde{S}}_{ij} \rangle. \tag{56}$$

The total Reynolds stress tensor is reconstructed by decomposing the total fluctuation into contributions from the resolved, deconvolved (soft-SFS), and modeled (hard-SFS) scales. This decomposition yields:

$$\langle u_i' u_j' \rangle \approx \langle (\bar{\tilde{u}}_i - \langle \bar{\tilde{u}}_i \rangle)\left(\bar{\tilde{u}}_j - \langle \bar{\tilde{u}}_j \rangle\right) \rangle + \langle b_{ij}^{uu} \rangle + \langle a_{ij}^{uu,r} \rangle + \frac{2}{3}\langle \overline{k_{\mathrm{SGS}}} \rangle \delta_{ij} \tag{57}$$

In Eq. (57), the first term on the right-hand side represents the resolved-scale Reynolds stress. The terms $\langle b_{ij}^{uu} \rangle$ and $\langle a_{ij}^{uu,r} \rangle$ are the mean contributions from the soft (deconvolved) and hard (modeled) SFS tensors, respectively. The final term is the isotropic component derived from the mean of the filtered SGS turbulent kinetic energy (TKE) $\langle \overline{k_{SGS}} \rangle = \langle \overline{k^{\star}} \rangle$. For the algebraic models considered in this study, the unfiltered $k_{SGS}$ is not computed from a transport equation but is instead estimated using the following scaling relationship (Sullivan *et al.* 2003):

$$\begin{aligned} k_{\mathrm{SGS}} &\approx \left(\frac{C_{\mathrm{D}}}{C_{\mathrm{e}}}\right)^{\frac{2}{3}} \tilde{\Delta}^2 S^{\star 2} \ ; (\mathrm{A-DEV}) \\ \overline{k_{\mathrm{SGS}}} &\approx \left(\frac{C_{\mathrm{D}}}{C_{\mathrm{e}}}\right)^{\frac{2}{3}} \bar{\tilde{\Delta}}^2 \bar{\tilde{S}}^2 \ ; (\mathrm{S-DEV}) \end{aligned} \tag{58}$$

where $C_e$ represents the model constant: $C_e = 1$ for LDME or $C_e = 1.048$ for DSM.

## 3. Benchmarks

### *3.1. The turbulent channel flow*

Turbulent flow in a plane channel, driven by a constant mean pressure gradient, serves as a canonical benchmark for *a priori* and *a posteriori* model validation. Its widespread usage is primarily due to the availability of extensive DNS databases. The flow dynamics is governed by a single dimensionless parameter, the friction Reynolds number, $Re_\tau = u_\tau H/\nu$. Here, $H$ is the half-channel height, $\nu$ denotes the kinematic viscosity, $u_\tau = \tau_w^{1/2}$ is the friction velocity, and $\tau_w$ is the wall shear stress. This investigation utilizes the case with $Re_\tau = 395$ of DNS dataset by Moser *et al.* (1999). The computational domain is configured with a cuboid box of dimension $(L \times 2H \times W) = (2\pi \times 2 \times \pi)$. Periodic boundary conditions are applied in the streamwise and spanwise directions for all flow variables, with the mean pressure gradient enforced through an additional source term in the momentum equation. At the lower and upper walls, the no-slip condition is imposed on the velocity field, while a zero-gradient condition is applied to the pressure. All statistics are presented as a function of the normalized wall-normal coordinate, $y^+ = yu_\tau/\nu$.

### *3.2. The turbulent two-phase mixing layer*

The turbulent two-phase mixing layer serves as the second canonical benchmark. This case is selected to assess model performance in a complex, non-equilibrium flow involving significant interfacial dynamics, such as shear-driven destabilization, ligament stretching, and spray formation. Its utility as a validation case is established by the availability of an extensive DNS database and detailed statistical analyses published by Ling et al. (2019). The statistical results presented in this work are validated against the DNS data derived from three complementary studies (Ling *et al.* 2017, Ling *et al.* 2019, Jiang & Ling 2020), all utilizing an identical physical configuration and numerical setup. This dataset allows for a multi-faceted assessment of the LES closures across momentum, scalar transport, and interfacial dynamics. The configuration, depicted in Figure 1, consists of a planar liquid stream and a parallel gas stream, initially separated by a thin plate of length $l_x$ and thickness $l_y$. The computational domain is a rectangular prism with dimensions $L_x \times L_y \times L_z = 16H_l \times 8H_l \times 2H_l$. The inlet liquid stream height, $H_l$, is used as the characteristic length scale. The coordinate system is defined with $x$ in the primary flow (streamwise) direction, $y$ in the transverse direction, and $z$ in the homogeneous (spanwise) direction. The simulation setup is based on the parameters from the reference DNS (Ling *et al.* 2019), summarized in Table 1.

At the inlet ($x = 0$), velocity profiles for the liquid and gas phases are defined using error functions based on the boundary layer thickness $\delta_g$ and the maximum velocities $U_l$ and $U_g$. The inlet liquid volume fraction $\alpha$ is set to 1 for $0 \leq y < H_l$ and 0 otherwise. A critical aspect of this benchmark, following the reference studies, is the use of laminar inflow conditions (zero inlet turbulence intensity). This ensures that all turbulence and interfacial instabilities are generated within the domain due to the inherent shear. Therefore, no synthetic turbulence generation (Wang *et al.* 2026) is required at inlet. The remaining boundary conditions are specified as follows: an outflow condition is applied at $x = L_x$, a free boundary at $y = L_y$, a slip wall at $y = 0$, and cyclic (periodic) conditions are imposed in the spanwise z-direction. The surfaces of the separator plate are defined as no-slip walls. Consistent with the reference DNS (Ling *et al.* 2019), an approximate convective outflow condition is employed.

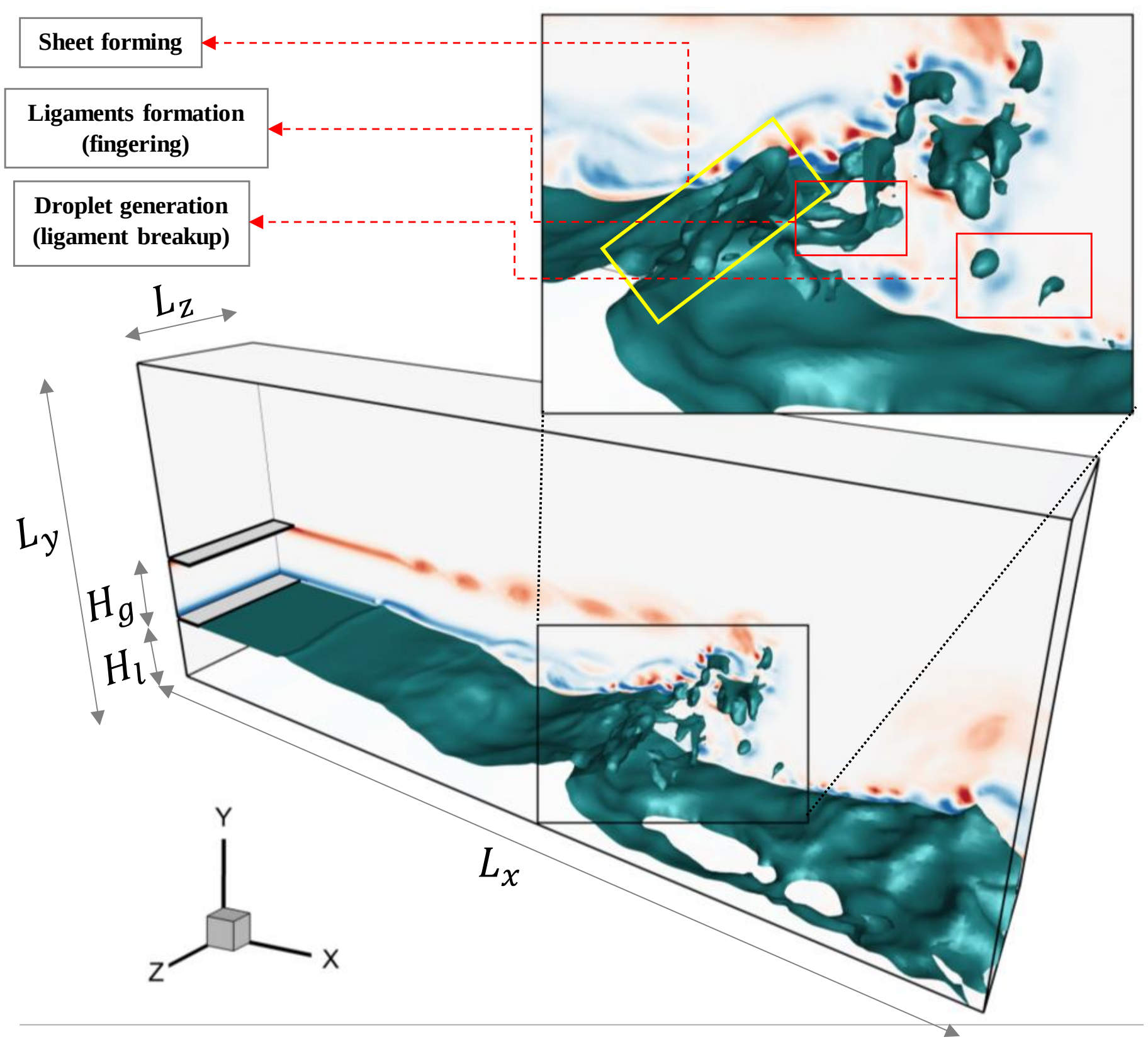

**Figure 1:** The two-phase mixing layer: instantaneous three-dimensional visualization of primary atomization and spray formation in the turbulent gas-liquid mixing layer. The intricate interfacial dynamics, including shear-driven Kelvin-Helmholtz instabilities and ligament ruptures, are predicted using the proposed one-equation mixed AD-LES framework (A-LDMK) augmented with the structural SFS surface tension closure (Run 5).

**Table 1:** Physical parameters for the two-phase mixing layer benchmark (Ling *et al.* 2019).

| $\rho_l$ $(\frac{kg}{m^3})$ | $\rho_g$ $(\frac{kg}{m^3})$ | $\mu_l$ $(Pa.s)$ | $\mu_g$ $(Pa.s)$ | $\sigma$ $(\frac{N}{m})$ | $U_l$ $(\frac{m}{s})$ | $U_g$ $(\frac{m}{s})$ | $H_l$ $(m)$ | $\delta_g$ $(m)$ | $l_x$ $(m)$ | $l_y$ $(m)$ |
|---|---|---|---|---|---|---|---|---|---|---|
| 1000 | 50 | $10^{-3}$ | $5 \times 10^{-5}$ | 0.05 | 0.5 | 10 | $8 \times 10^{-4}$ | $10^{-4}$ | $4 \times 10^{-5}$ | $2.5 \times 10^{-5}$ |

## 4. Numerical method

In this study, the cell-centered collocated Finite Volume Method (FVM) implemented in the OpenFOAM CFD package, version 1912 (www.openfoam.com), was extended to incorporate ADM-type closures.

For the explicit filtering of width $\bar{\Delta}= 2\tilde{\Delta}= 2h$ or the box filtering of width $\breve{\Delta}= 2\tilde{\Delta}= 2h$, required for the A-LDME and A-LDMK models in the formulation given in section 2.2.2, the "simpleFilter" is utilized, which is defined as:

$$\bar{\phi}_i = \frac{\sum_{f \in \text{faces}} \phi_{f,i} |\mathbb{S}_{f,i}|}{\sum_{f \in \text{faces}} |\mathbb{S}_{f,i}|}, \tag{59}$$

where $|\mathbb{S}_{f,i}|$ denotes the area of face $f$ of computational cell $i$, $\phi_{f,i}$ is the interpolated value of the variable $\phi$ at the face $f$ centroid. Note that $\tilde{\Delta}$ is chosen as $(V_{\text{cell}})^{1/3}$. This filtering operation computes the entire filtered field through a two-step process: interpolation from cell centers to faces, followed by a surface-averaging step back to the cell centers. For the larger explicit filtering of width $\hat{\Delta}= 2\bar{\Delta}= 4h$, required for the S-LDME model described in section 2.2.1, this recursive operation is executed five consecutive times. The requisite number of iterations is determined by enforcing the condition $\hat{G}(\pi/2) = 0.5$ or $\hat{G}(\pi/4) = 0.5$ (see Figure 2) for the filter width of $2h$ or $4h$, respectively (Vasilyev *et al.* 1998, Stolz *et al.* 2001). The transfer function of this operator, depicted in Figure 2 (for different number of recursions), exhibits fundamental properties essential for explicit filtering, such as normalization ($\hat{G}(\omega' = 0) = 1$), positivity, smoothness, and zero response at the Nyquist limit ($\hat{G}(\pi) = 0$) (Vasilyev *et al.* 1998, Najafiyazdi *et al.* 2023).

During the computational campaign, numerical divergence was encountered when employing mixed AD-DEV formulations. Initially, a clipping procedure rooted in the realizability conditions proposed by Mokhtarpoor and Heinz (2017) for pure DEV-LES was applied; however, this proved insufficient for the present AD-LES to stabilize the solution across multiple test cases. To resolve these instabilities, a Positive Total Viscosity (PTV) constraint ($\nu_t + \nu \geq 0$ or $\nu_t/\nu \geq -1$) was enforced for some mixed AD-DEV simulations diverged without PTV, which will be identified later. A detailed analysis of these stability challenges and the detailed impact of the PTV approach are thoroughly discussed in the results and discussion section. Other aspects of the numerical method are specific to each benchmark cases, which are described in the subsequent sub-sections.

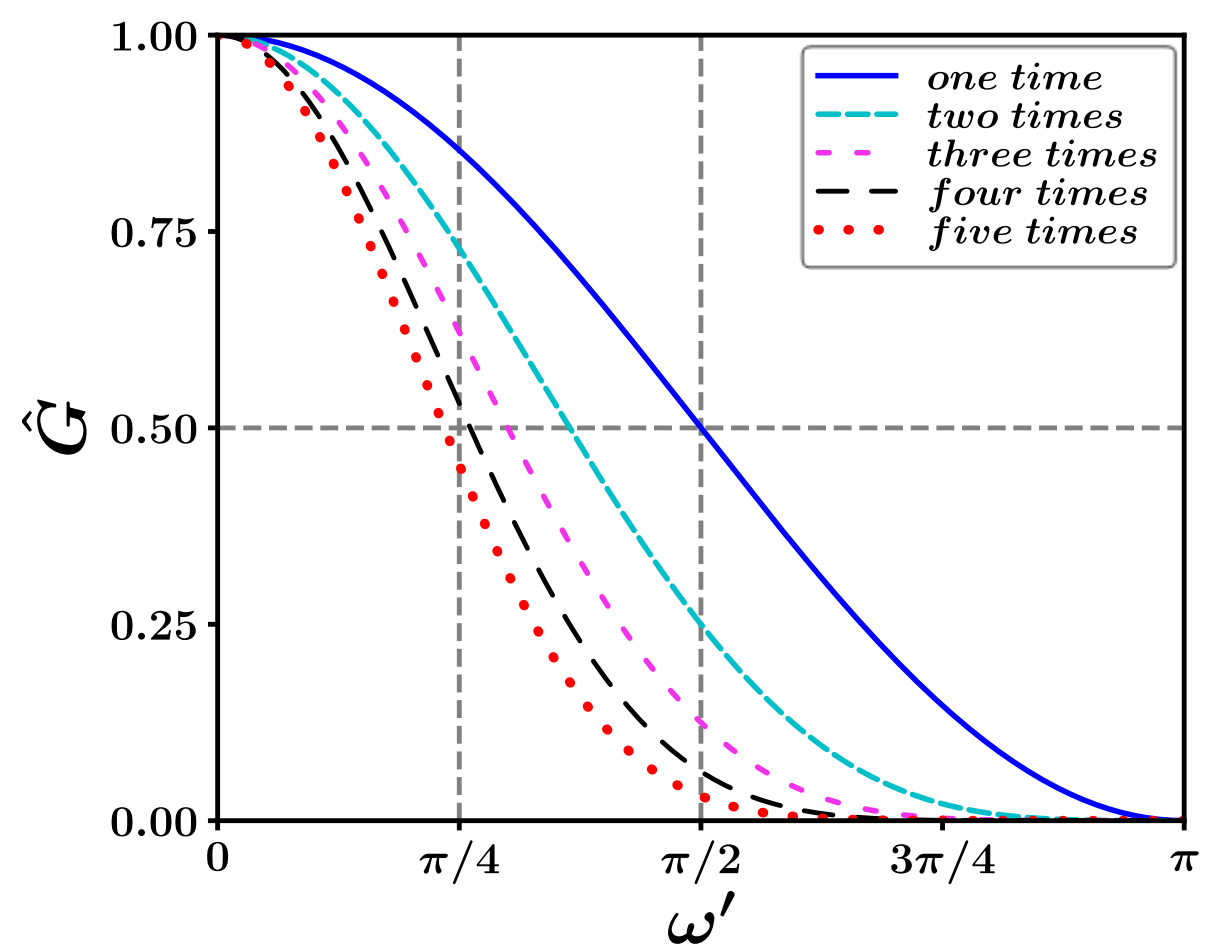

**Figure 2:** The recursive filter transfer function for different numbers of filtering operations.

### *4.1. The turbulent channel flow*

For the present LES of channel flow, the pressure-velocity coupling is resolved using the Pressure Implicit with Splitting of Operators (PISO) algorithm (Issa 1986) employing two pressure-correction loops. Temporal discretization is achieved using the implicit second-order backward scheme (Greenshields 2015, Moukalled et al. 2016). Gradients are computed via the Green-Gauss cell-based method, utilizing the "Gauss linear" scheme (Greenshields 2015). The advection term in the momentum and TKE equations are discretized using the second-order pure-centered "Gauss linear" scheme, while variable interpolation to cell faces is performed using the "linear" scheme. For the pressure equation, the resulting system of linear algebraic equations is solved using the Preconditioned Conjugate Gradient (PCG) method, with the Diagonal Incomplete-Cholesky (DIC) preconditioner combined with Gauss-Seidel smoothing (Saad 2003). The momentum equation is solved using the Stabilized Preconditioned Bi-Conjugate Gradient (PBiCGStab) solver, preconditioned with the Diagonal Incomplete-LU (DILU) method (Van der Vorst 1992, Barrett et al. 1994, Saad 2003). A normalized residual tolerance of $10^{-6}$ is set as the convergence criterion for all variables at each time step. To ensure numerical stability and accuracy, the time step is dynamically adjusted to maintain a maximum Courant-Friedrichs-Lewy (CFL) number of 0.5. The computational grid is uniform in the streamwise and spanwise directions, with wall-normal refinement achieved using a hyperbolic tangent function to cluster points near the walls. Details of the LES grid configurations, along with the reference DNS grid, are provided in Table 2.

**Table 2:** A comparison of the present LES and reference DNS (Moser et al. 1999) grids.

| | $\Delta x^+$ | $\Delta y^+$ | $\Delta z^+$ | $\widetilde{\Delta}/h_{DNS}$ | $N_x \times N_y \times N_z$ |
|---|---|---|---|---|---|
| **DNS** | 10 | [0.03, 6.5] | 6.5 | 1 | $256 \times 193 \times 192 = 9486336$ |
| **LES (Coarse)** | 38.78 | [0.82, 25] | 25.85 | 3.04 | $64 \times 110 \times 48 = 337920$ |
| **LES (Fine)** | 50.65 | [0.82, 25] | 15.13 | 2.78 | $49 \times 110 \times 82 = 441980$ |

Owing to its simplicity and inherent robustness, the Wall-Adapting Local Eddy-viscosity (WALE) model (Nicoud & Ducros 1999) is employed to generate the initial flow fields, thereby accelerating the convergence of the advanced models. The details regarding the velocity field initialization and the computational procedure for the WALE precursor simulation are available in prior studies (Amani *et al.* 2023, Taghvaei & Amani 2023). Initiated from this WALE baseline, the primary simulations are advanced for 30 Flow-Through-Times (FTT $= L/U_b$, where $U_b$ is the channel bulk velocity) to ensure the flow reaches a statistically stationary state. Once this stationary regime is established, data is collected over an additional 70 FTT. Flow statistics, denoted by the ensemble average $\langle \quad \rangle$, are compiled through time-averaging combined with spatial averaging across the homogeneous streamwise ($x$) and spanwise ($z$) directions, after elapsing the initial transient.

### *4.2. The turbulent two-phase mixing layer*

For the simulation of the turbulent two-phase mixing layer, the pressure-velocity coupling was managed by the PISO algorithm, which was configured with three inner pressure-correction loops. Similar to the channel flow, gradient computation utilized the Gauss linear scheme, and variable interpolation was performed using the linear scheme. The advection schemes, however, were specialized for this multiphase solver: the momentum advection term was discretized using the second-order, bounded Gauss linear scheme. To maintain a sharp interface, the current discretization of the VOF transport equation involves the Gauss vanLeer scheme for phase advection and the algebraic Multidimensional Universal Limiter with Explicit Solution (MULES) flux limiter (Kuzmin *et al.* 2003).

The linear solver for the relative pressure was, consistent with the previous case, the PCG method with a DIC preconditioner. The momentum and turbulence transport equations, conversely, were solved using a smoothSolver based on the symmetric Gauss-Seidel smoother. A stricter normalized residual tolerance of $10^{-9}$ was set for all variables. Temporal discretization was achieved using the first-order implicit Euler scheme. Numerical stability and accuracy were ensured via an adaptive time step, maintaining a maximum CFL number of 0.1 for the volume fraction equation and 0.25 for the other equations.

To optimize computational efficiency while preserving the essential flow physics, the LES computational domain is slightly truncated compared to the reference DNS configuration, spanning $L_x \times L_y \times L_z = 12H_l \times 5H_l \times 2H_l$. This reduction in the streamwise and transverse directions targets the far-field regions where the influence on the primary atomization dynamics is minimal. The computational domain is discretized using a Cartesian cut-cell grid featuring multi-level local refinement to resolve multiscale flow structures, as depicted in Figure 3. To accurately capture the boundary layer characteristics and initial shear layer instabilities, the highest grid resolution is concentrated near the separator plates (zone 1, 2, and 3), ensuring a dimensionless wall distance of $y^+ \approx 1$. A series of buffer layers (zone 4) is employed to facilitate a smooth transition from the near-wall regions to the core mixing zone (zone 5), where primary atomization occurs. Detailed specifications of the grid resolution for each identified topological zone are summarized in Table 3. The total number of grid cells in our LES setup is about 1.2M, compared to 537M cells used in the reference DNS.

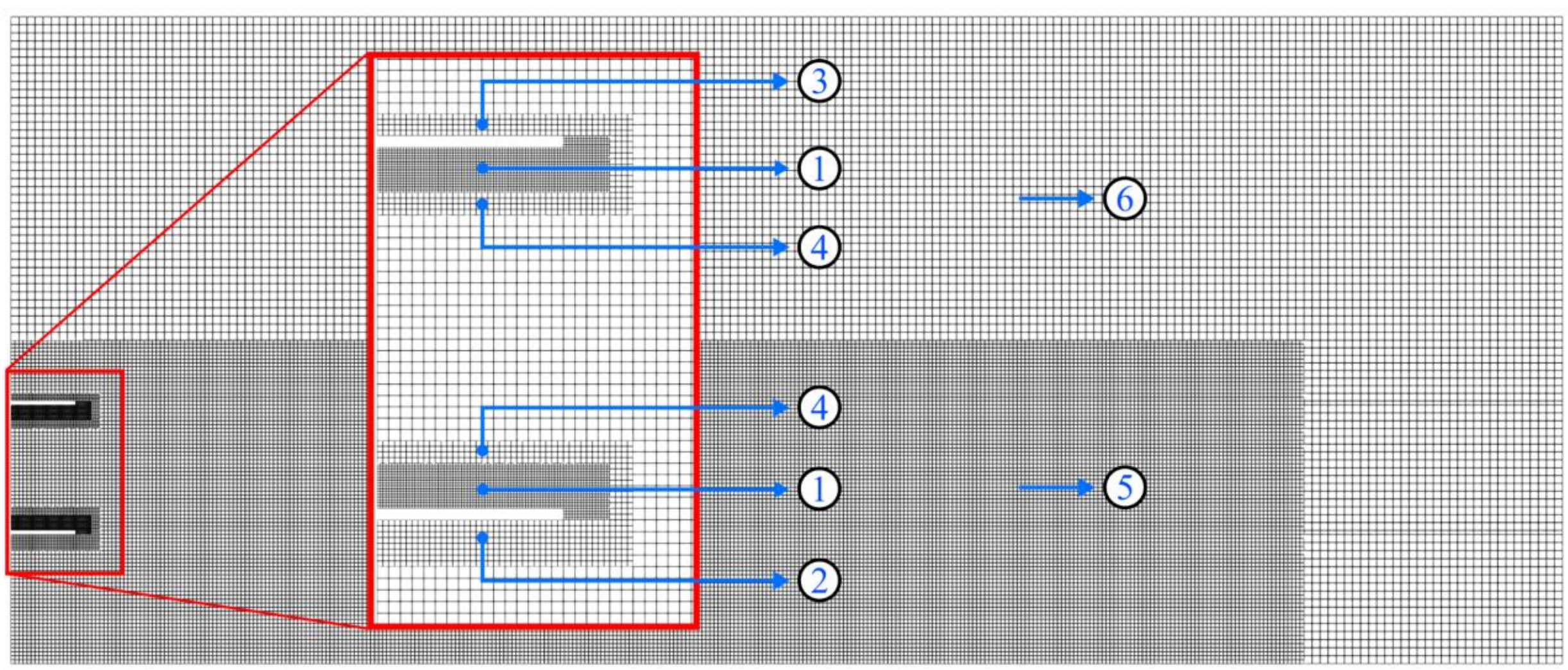

**Figure 3:** The computational grid topology for the two-phase mixing layer. The red inset highlights the multi-level cut-cell refinement strategy near the separator plates, including the boundary layers and transition zones. Detailed grid resolutions for the labeled regions (1-6) are provided in Table 3.

**Table 3:** Grid resolution specifications for the two-phase turbulent mixing layer LES.

| | Name | $\Delta x$ ($\mu m$) | $\Delta y$ ($\mu m$) | $\Delta z$ ($\mu m$) | $\widetilde{\Delta}$ ($\mu m$) | $\widetilde{\Delta}/h_{DNS}$ (-) |
|---|---|---|---|---|---|---|
| **1** | Gas near-wall boundary layers | 6.25 | 6.25 | 50 | 12.5 | 4 |
| **2** | Liquid near-wall boundary layers | 12.5 | 12.5 | 50 | 19.84 | **6.34** |
| **3** | Low-velocity gas boundary layers | 12.5 | 12.5 | 50 | 19.84 | **6.34** |
| **4** | Buffer layers (mesh transition) | 12.5 | 12.5 | 50 | 19.84 | **6.34** |
| **5** | Core mixing region (primary atomization) | 25 | 25 | 50 | 31.50 | 10.16 |
| **6** | Far-field domain | 50 | 50 | 50 | 50 | 16 |

To quantitatively assess the adequacy of the computational grid, the Pope's criterion was evaluated to measure the ratio of the resolved-to-total turbulent kinetic energy. Figure 4 presents the transverse profiles of this ratio at three representative locations. The results demonstrate that the criterion remains above the recommended 80% threshold across the vast majority of the computational domain. Although the criterion drops slightly below 80% (reaching approximately 75%) in very narrow, highly localized regions within the high-speed gas jet, given our focus on comparing SGS closures, operating near the lower margin of a well-resolved LES (80% kinetic-energy resolution) is favorable because it provides a sufficiently large SGS contribution for a meaningful comparison, consistent with the conclusion for the single-phase-flow benchmark, where the coarse grid, rather than the fine grid shown in Figure 5a, was selected for analyses.

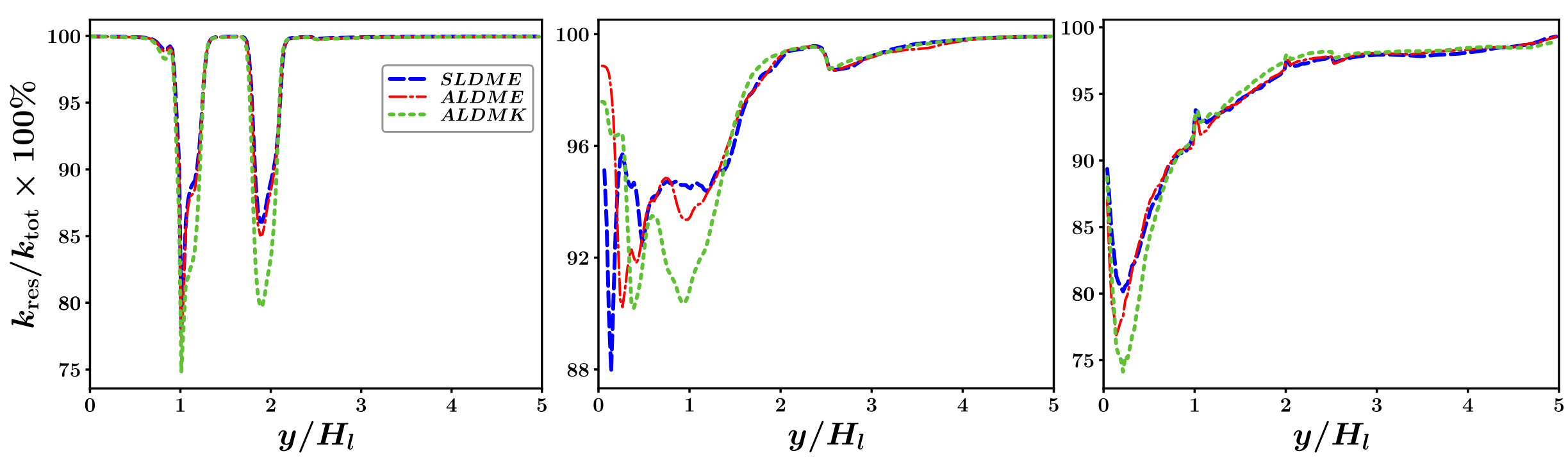

**(a)** **(b)** **(c)**

**Figure 4**: The "turbulent two-phase mixing layer" benchmark: Transverse profiles of the ratio of the resolved-to-total turbulent kinetic energy, at three cross-sections: (a) $x = 2H_l$, (b) $x = 6H_l$, and (c) $x = 10H_l$ for different AD-LES closures.

The simulations are initialized with a quiescent gas phase filling the entire computational domain. Following the injection of the gas and liquid streams, a transient start-up period is observed. Consistent with the reference DNS study by Ling et al. (2017) monitoring the domain-integrated kinetic energy indicates that statistically stationary state is established after approximately $t = 200H_l/U_g \approx 16$ ms. Consequently, to ensure statistical convergence, the simulations are extended to a final physical time of $t > 70$ ms, with data sampling and time-averaging performed over the post-transient period ($t > 16$ ms). Flow statistics are computed by averaging in the homogeneous spanwise direction ($z$) and in time.

## 5. Results and discussion

A comparative analysis of algebraic S-DEV and A-DEV closures for a single-phase turbulent channel flow was conducted by Amani *et al.* (2024). While the models produced similar results for primary statistics on fine grids, the S-DEV model showed a near-zero contribution from the modeled stress $a_{ij}^{uu}$. This diminished contribution was compensated by the deconvolved term $b_{ij}^{uu}$, a behavior that became problematic on coarser grids. The A-DEV model operated without the need for clipping, whereas the S-DEV model required clipping to maintain stability. In Section 5.1, we assess the performance of the proposed one-equation AD-LES model in the single-phase turbulent channel flow benchmark. Then, in Section 5.2, this family of AD-LES models, extended to a two-phase flow formulation, is evaluated and compared to standard closures in the two-phase mixing layer benchmark.

### *5.1.The turbulent channel flow*

The performance of the newly proposed one-equation mixed AD-LES model (A-LDMK) is evaluated against the reference DNS data of Moser *et al.* (1999), in comparison to the algebraic counterpart of the same consistent mixed AD-LES family of models (A-LDME), the standard mixed AD-LES (S-LDME), and the conventional WALE pure EV-LES. Figure 5 illustrates the profiles of the mean streamwise velocity and the Reynolds stress tensor components. To establish a rigorous baseline for numerical accuracy and evaluate grid sensitivity, the mean streamwise velocity profiles (Figure 5a) are examined for both the fine and coarse grid configurations detailed in Table 2. The consistency of the solver and approaching to DNS solution by reducing the grid resolution and filter width is evident from this figure. When utilizing the fine grid (and filter) resolution, all evaluated mixed AD-LES predictions, namely A-LDMK, A-LDME, and S-LDME, are very close to each other, successfully capturing the correct asymptotic DNS behavior and showing slight improvements compared to pure WALE EV-LES. However, to evaluate the SGS closures under more challenging conditions and observe the differences in their predictive capabilities, the coarser grid setup is selected for the remaining investigations.

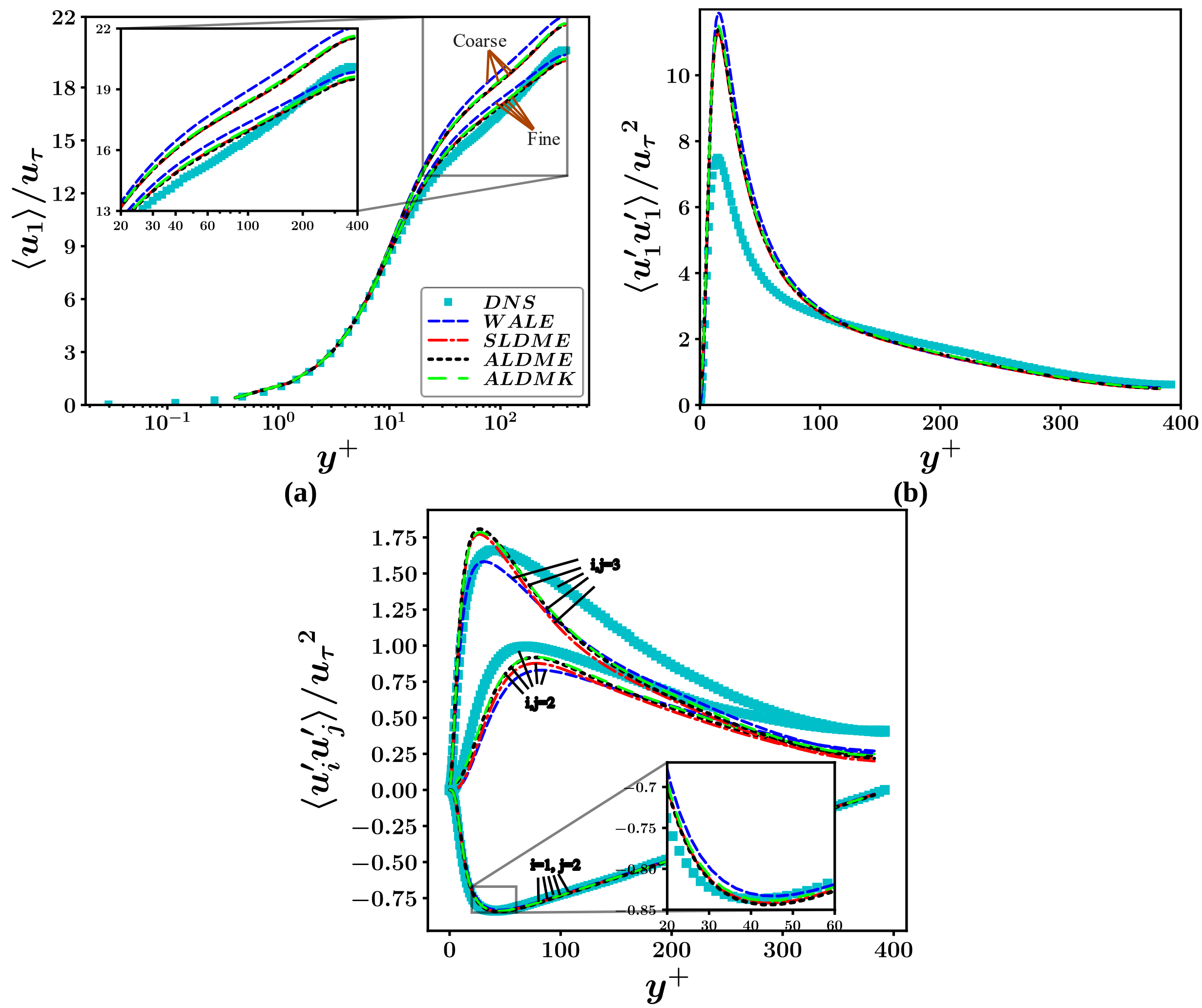

**Figure 5:** The "channel flow" benchmark: The comparison of different LES closures against the DNS data (Moser *et al.* 1999) for the mean streamwise velocity (a) (the fine and coarse grids) and total Reynolds stress components (the course grid): streamwise (b), wall-normal, spanwise, and shear (c).

Accordingly, unless otherwise specified, all subsequent turbulence statistics and profiles presented in this section are compiled strictly from the coarse grid simulations. As observed in the coarse grid results in Figure 5a, the standalone WALE model shows more deviations beyond the buffer layer, while the mixed AD-LES closures do not resolve the log-layer mismatch on the coarse grid, they provide a closer agreement with the DNS data in the log-law region. The deviation of the WALE model from the DNS data is also visible in the Reynolds stress components, with the exception of the spanwise fluctuations (see Figure 5b and c). The mixed models yield comparable predictions of the mean velocity, as well as the streamwise normal (Figure 5b) and shear (Figure 5c) Reynolds stresses.

To quantitatively assess the performance of the evaluated closures, the error norm, $EN(Q)$, for each statistical quantity $Q$ is computed by

$$EN(Q) = \frac{\mathcal{N}(Q_{\mathrm{LES}} - Q_{\mathrm{FDNS}})}{\mathcal{N}(Q_{\mathrm{FDNS}})}, \tag{60}$$

where the norm operator, $\mathcal{N}(.)$, represents the spatial and temporal $L_2$-norm, defined by

$$\mathcal{N}(Q) = \langle Q^2 \rangle^{1/2}. \tag{61}$$

The computed error norms for the coarse grid simulations are summarized in Table 4. This quantitative analysis corroborates the visual profile comparisons. Among the AD-LES closures, S-LDME yields the lowest errors for the mean velocity, $\langle u_1 \rangle$, and the streamwise and shear stresses, $\langle u_1' u_1' \rangle$ and $\langle u_1' u_2' \rangle$, however, the differences between S-LDME, A-LDME, and A-LDMK for these specific variables are marginal, below 0.5%, 1.7%, 1.9%, and 0.3% for $\langle u_1 \rangle$, $\langle u_1' u_1' \rangle$, $\langle u_3' u_3' \rangle$, and $\langle u_1' u_2' \rangle$, respectively. The major distinction is observed in the prediction of the wall-normal fluctuations, $\langle u_2' u_2' \rangle$, which serve as a primary discriminator among the mixed models. Specifically, the error norm analysis shows that the S-LDME formulation yields a higher error for $\langle u_2' u_2' \rangle$ with a value of 24.2%, which is comparable to the WALE model error (25.8%). On the other hand, A-LDMK exhibits 7.1% and 8.7% error reductions compared to S-LDME and WALE results, respectively.

Generally, among all closures, the standalone WALE model resulted in the largest errors for all statistics with the exception of the spanwise fluctuations $\langle u_3' u_3' \rangle$, where WALE yielded a 3.2% lower error than A-LDMK. The highly dissipative nature of the pure WALE model artificially damps the turbulent fluctuations, which coincidentally aligns the spanwise component closer to the reference data on this coarse grid. The AD-LES models retain more resolved energy, leading to a slight overprediction in the spanwise direction, although the consistent A-LDMK and A-LDME models improve this prediction relative to S-LDME.

**Table 4:** The "channel flow" benchmark: The error norm of the predicted statistics by different AD-LES approaches on the coarse LES grid introduced in Table 2.

| Grid | Statistics | Error norm | | | |
|---|---|---|---|---|---|
| | | WALE | SLDME | ALDME | ALDMK |
| Coarse ($\tilde{\Delta}/h_{DNS} \sim 3.04$) | $\langle u_1 \rangle$ | 0.121 | 0.092 | 0.094 | 0.097 |
| | $\langle u_1' u_1' \rangle$ | 0.389 | 0.345 | 0.354 | 0.362 |
| | $\langle u_2' u_2' \rangle$ | 0.258 | 0.242 | 0.185 | 0.171 |
| | $\langle u_3' u_3' \rangle$ | 0.117 | 0.168 | 0.162 | 0.149 |
| | $\langle u_1' u_2' \rangle$ | 0.045 | 0.038 | 0.041 | 0.040 |

An observation is that the standard S-LDME predicts primary statistics with reasonable accuracy, despite being C1-inconsistent. To show this inconsistency *a posteriori*, the contributions of the deconvolved ($b_{ij}^{uu}$) and modeled ($a_{ij}^{uu}$) stresses are isolated in Figure 6. As shown in Figure 6b and d (blue dashed lines), the modeled SGS stress components of S-LDME are near zero. This indicates that the standard procedure ignores the modeled contribution ($a_{ij}^{uu}$) when used alongside deconvolution. Consequently, the prediction of the total stress by S-LDME appears to rely on a compensation effect: the deconvolved part increases in magnitude (see Figure 6a and c) to balance the vanishing modeled part. In contrast, both A-LDMK (black dashed lines) and the algebraic A-LDME model (red dash-dotted lines) maintain a physically consistent behavior, with contributions originating from both the soft and hard deconvolution terms, confirming the efficacy of the consistent closure.

Regarding numerical stability, the models exhibit different characteristics. Figure 7 displays the profiles of both the time-averaged (Figure 7a) and the instantaneous (Figure 7b) SGS eddy viscosity ratio, $\langle \nu_r/\nu \rangle$. Consistent with the stress analysis, the S-LDME yields a near-zero mean eddy viscosity, whereas the A-LDME and A-LDMK models produce finite viscosity levels. Standard functional models, such as the WALE formulation, impose a positive eddy viscosity and do not account for energy backscatter (Park & Mahesh 2008). While subgrid models must provide adequate forward dissipation in expectation (Moser *et al.* 2021), the subgrid stress can exhibit instantaneous backward dissipation. Studies by Piomelli *et al.* (1991) have shown that kinetic energy backscatter from the subgrid to the resolved scales occurs in turbulent flows. Capturing these local backscatter events, while maintaining global forward-scatter, is a useful feature for SGS models (Park & Mahesh 2008). In the formulations presented here, the instantaneous viscosity can attain negative values, allowing the models to represent backscatter events. However, accommodating backscatter can introduce numerical instabilities (Park & Mahesh 2008). To maintain stability, the S-LDME model required the implementation of the PTV constraint ($\nu_r/\nu \geq -1$). In contrast, the A-LDMK and algebraic A-LDME model remained stable for the channel flow benchmark without requiring a clipping or a PTV constraint, accommodating instantaneous viscosity ratios as low as $\nu_r/\nu \approx -4$ (Figure 7b). It should be noted that while A-LDMK needed no clipping and PTV constraint on $\nu_r$ in the momentum equation, the diffusion term in the TKE transport equation, Eq. (55), should be treated with special care in such a manner that the diffusion coefficient in this equation, i.e., $\mu + \mu_r$, was clipped to prevent from becoming negative, otherwise divergence occurred. This is equivalent to a PTV constraint ($\nu_r \geq -\nu$) used on the TKE transport equation rather than the momentum equation.

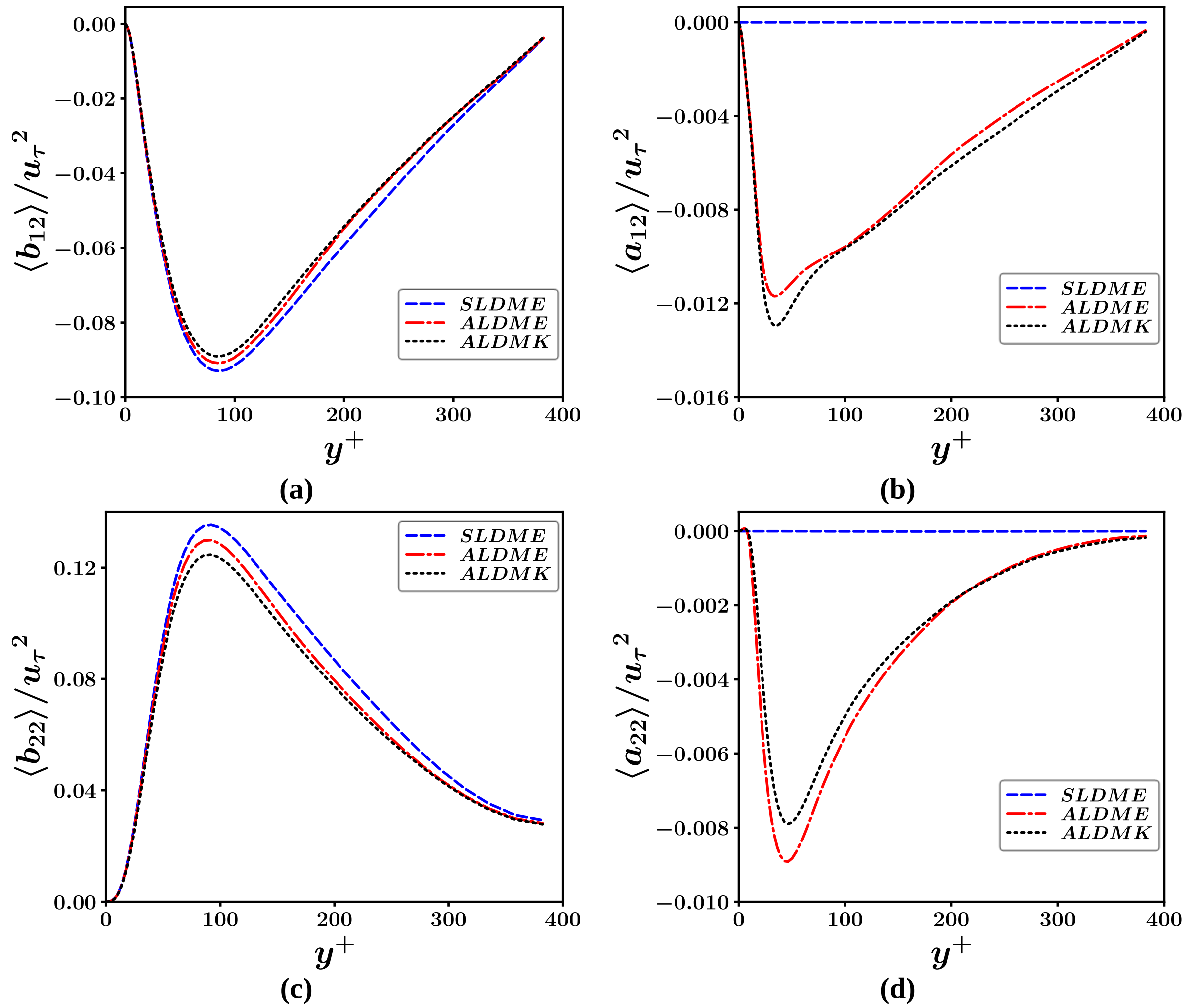

**Figure 6:** The “channel flow” benchmark: The comparison of the deconvolved, $b_{ij}$, and modeled, $a_{ij}$, SFS stress components profiles using different mixed AD-LES approaches.

In summary, the A-LDME and A-LDMK models offer performance improvements in terms of accuracy and solution stability compared to the S-LDME closure. The A-LDME framework demonstrates improved robustness in this configuration. Additionally, it provides higher computational efficiency by avoiding the need to solve an additional transport equation for the SGS kinetic energy and by operating without the reliance on auxiliary stabilization models.

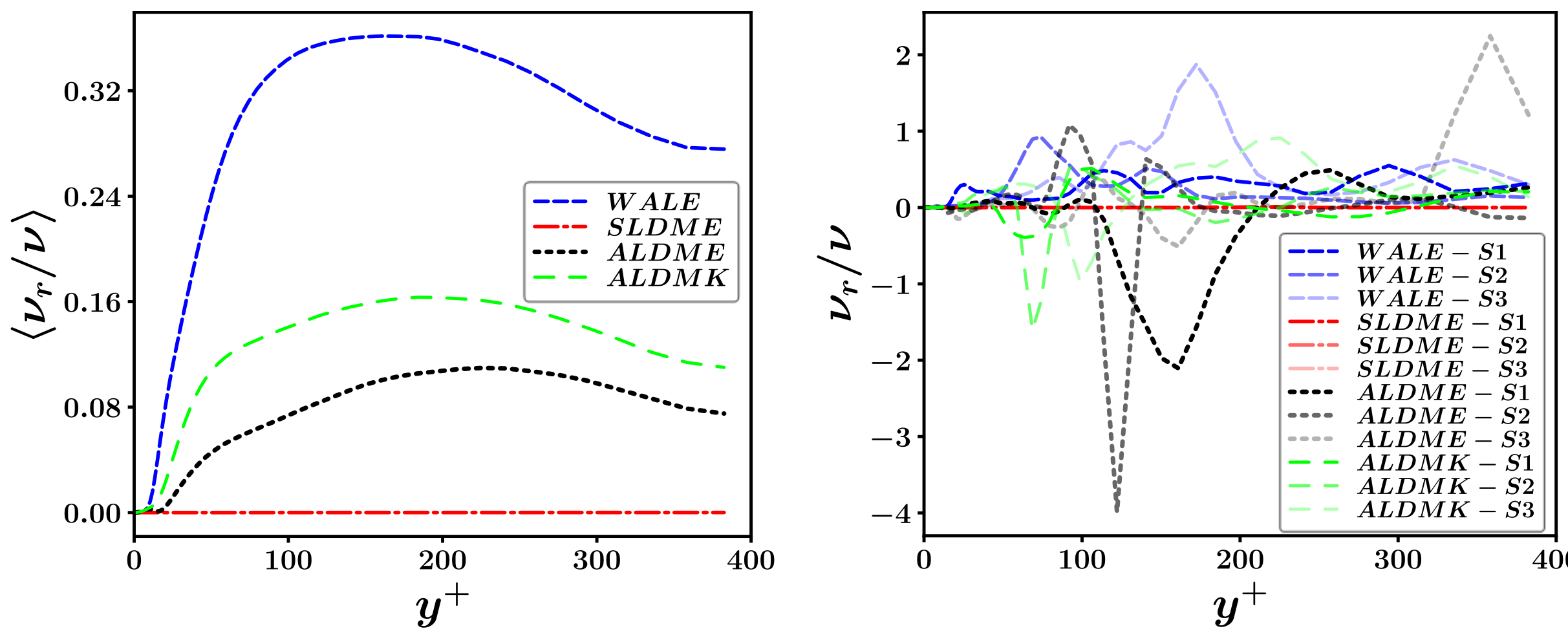

**(a)** **(b)**

**Figure 7**: The "channel flow" benchmark: A comparison of the (normalized) SGS eddy-viscosity using different LES approaches. The averaged (a) and three samples of instantaneous (b) profiles.

### *5.2. The turbulent two-phase mixing layer*

A visualization of the present AD-LES solution for the two-phase mixing layer benchmark using the A-LDMK closure is illustrated in Figure 1. This representation captures the complex physical cascade of primary atomization. The transition initiates with the development of shear-driven Kelvin-Helmholtz instabilities at the gas-liquid interface, concurrent with vortex shedding in the upper gas-gas mixing layer. As the interfacial waves grow and roll up, they stretch into thin liquid structures, triggering sheet thinning and flapping dynamics. Subsequently, transverse instabilities (e.g., Rayleigh-Taylor) manifest at the leading edges of these sheets, leading to the formation of longitudinal ligaments (fingering). The cascade culminates in the rupture of these ligaments (via Rayleigh-Plateau instabilities), governing the spray formation and the final droplets size distribution characteristics.

To systematically evaluate the performance of the new mixed AD-LES methodologies and to isolate the individual contributions of specific SFS terms, a detailed test matrix comprising five distinct simulation scenarios was established, as detailed in Table 5. The investigation is structured into two primary phases. First, the efficacy of the momentum closures is assessed by comparing the consistent one-equation model (A-LDMK) against the algebraic consistent formulation (A-LDME), the standard dynamic approach (S-LDME), and the conventional WALE model serving as a baseline. Based on the literature review, the most important closure is the SGS tensor ($\tau_{ij}^{uu}$), the SFS inertia tensor $T_{ij}^{uu} = b_{ij}^{uu} + a_{ij}^{uu}$. In all setups, SSADM, Eq. (31), is employed for the deconvolved part of the inertia tensor, $b_{ij}^{uu}$. The first three AD-LES modeling setups (Runs 1–3), respectively, involve A-LDMK, Eqs. (42)-(50) and (53)-(55), A-LDME, Eqs. (42)-(50) and (52), and S-LDME, Eqs. (36)-(39) and (41), for the modeled part of the inertia tensor, $a_{ij}^{uu}$, while neglecting SFS surface tension ($T_i^{u\sigma} = b_i^{u\sigma} + a_i^{u\sigma}$), and SFS phase transport ($T_i^{\alpha u} = b_i^{\alpha u} + a_i^{\alpha u}$). Second, to elucidate the physical impact of unresolved capillary forces, the structural reconstruction of the deconvolved part of the surface tension term ($b_i^{u\sigma}$), using SSADM, Eq. (32), is explicitly activated in Runs 4 and 5. It is pertinent to note that an attempt was made to further incorporate the deconvolved phase transport term ($b_i^{\alpha u}$ using SSADM, Eq. (33),) in Runs 6–8; however, this configuration proved numerically unstable, ultimately leading to divergence. This instability was found to be due to unphysical numerical oscillations of the volume fraction near the wall boundaries, when $b_i^{\alpha u}$ is included. Since we focus on subgrid momentum and surface tension closures, $b_i^{\alpha u}$ is omitted here, leaving the resolution of this divergence issue to future work. Consequently, the subsequent statistical analyses and discussions are strictly confined to the numerically stable configurations (Runs 1–5) where the effect of SFS phase transport is neglected.

**Table 5:** Test matrix for the turbulent two-phase mixing layer simulations, detailing the SFS modeling strategies.

| Case | Tag | $b_{ij}^{uu}$ | $a_{ij}^{uu}$ | $b_i^{u\sigma}$ | $b_i^{\alpha u}$ | Convergency |
|---|---|---|---|---|---|---|
| **1** | ALDMK | SSADM | A-LDMK | 0 | 0 | Converged |
| **2** | ALDME | SSADM | A-LDME | 0 | 0 | Converged |
| **3** | SLDME | SSADM | S-LDME | 0 | 0 | Converged |
| **4** | ALDME+SFST | SSADM | A-LDME | SSADM | 0 | Converged |
| **5** | ALDMK+SFST | SSADM | A-LDMK | SSADM | 0 | Converged |
| **6** | ALDMK+SFPT | SSADM | A-LDMK | 0 | SSADM | Diverged |
| **7** | ALDME+SFPT | SSADM | A-LDME | 0 | SSADM | Diverged |
| **8** | SLDME+SFPT | SSADM | S-LDME | 0 | SSADM | Diverged |

In terms of robustness, the highly transient nature of the turbulent two-phase mixing layer—characterized by intense interfacial shear and sharp density gradients—imposes rigorous demands on numerical stability. In the preceding single-phase channel flow benchmark, the consistent algebraic framework (A-LDME) exhibited improved robustness, remaining stable without any *ad hoc* clipping in this setup. However, the severe non-linearities associated with macroscopic interface deformations and primary atomization necessitate explicit numerical stabilization. Consequently, to accommodate physical kinetic energy backscatter without inducing solver divergence, both the new A-LDMK and A-LDME frameworks required the implementation of the standard PTV constraint ($\nu_r \geq -\nu$).

In contrast, the standard dynamic formulation (S-LDME) demonstrated lower numerical stability. For the S-LDME simulations, the standard PTV condition proved entirely insufficient to prevent divergence. To maintain a stable solution, the standard model necessitated a markedly more restrictive *ad hoc* clipping threshold ($\nu_r \geq -0.5\nu$). This degradation in model robustness could be linked to the vanishing modeled SFS stress in S-LDME, which could have served as a mechanism for energy removal from the flow, i.e., the CI-inconsistency of S-LDME shown for the channel flow benchmark and will also be revisited in this section for the two-phase flow case. This highlights an advantage of the A-LDME and A-LDMK methodologies: they allow for a broader representation of local backscatter, whereas the S-LDME model requires stabilization that can suppress certain flow dynamics.

Figure 8 illustrates the spatial evolution of the normalized mean streamwise velocity, $\langle u_1 \rangle / U_g$, across various longitudinal and transverse sections. This analysis evaluates the predictive capabilities of the assessed SGS closures in capturing the shear layer development, momentum diffusion, and cross-phase interactions. In the immediate vicinity of the inlet and the early mixing region, specifically at $x = 0.5H_l$ (Figure 8a) and $x = 2.0H_l$ (Figure 8b), the shear layer is highly confined, and turbulence is not yet fully developed. In these regions, all LES frameworks demonstrate fine agreement with the DNS data, accurately capturing the sharp velocity gradient across the gas-liquid interface. The influence of the structural Sub-Filter Surface Tension (SFST) model is understandably negligible in this early stage, as macroscopic interfacial deformations and primary droplet formations have not yet manifested.

The transverse profiles further downstream at $x = 6.0H_l$ (Figure 8d) clearly highlight the distinctions among the momentum closures. As the mixing layer expands, focusing on the inner shear layer and the high-speed gas stream region ($y/H_l < 1.5$), the standalone WALE model underpredicts the mean velocity due to its dissipative nature. Conversely, the A-LDMK and A-LDME frameworks demonstrate closer agreement with the DNS benchmark in this critical zone.

Furthermore, the inclusion of SFST closure yields relatively low levels of modulation in the mean velocity field for both the A-LDME and A-LDMK frameworks. As observed in the transverse profiles (Figure 8c and d), accounting for the SFST term produces only marginal deviations from the baseline models. This indicates that, for the present case, the macroscopic velocity distribution is predominantly dictated by the core momentum closures, with the sub-grid capillary forces exerting a negligible influence on the mean momentum field at this resolution.

While the consistent models successfully rectify the artificial dissipation inherent to standard EV closures in the inner shear layer, it is observed that near the global velocity peak ($y/H_l \approx 1.5$), the collective LES frameworks slightly overpredict the retained momentum compared to the DNS.

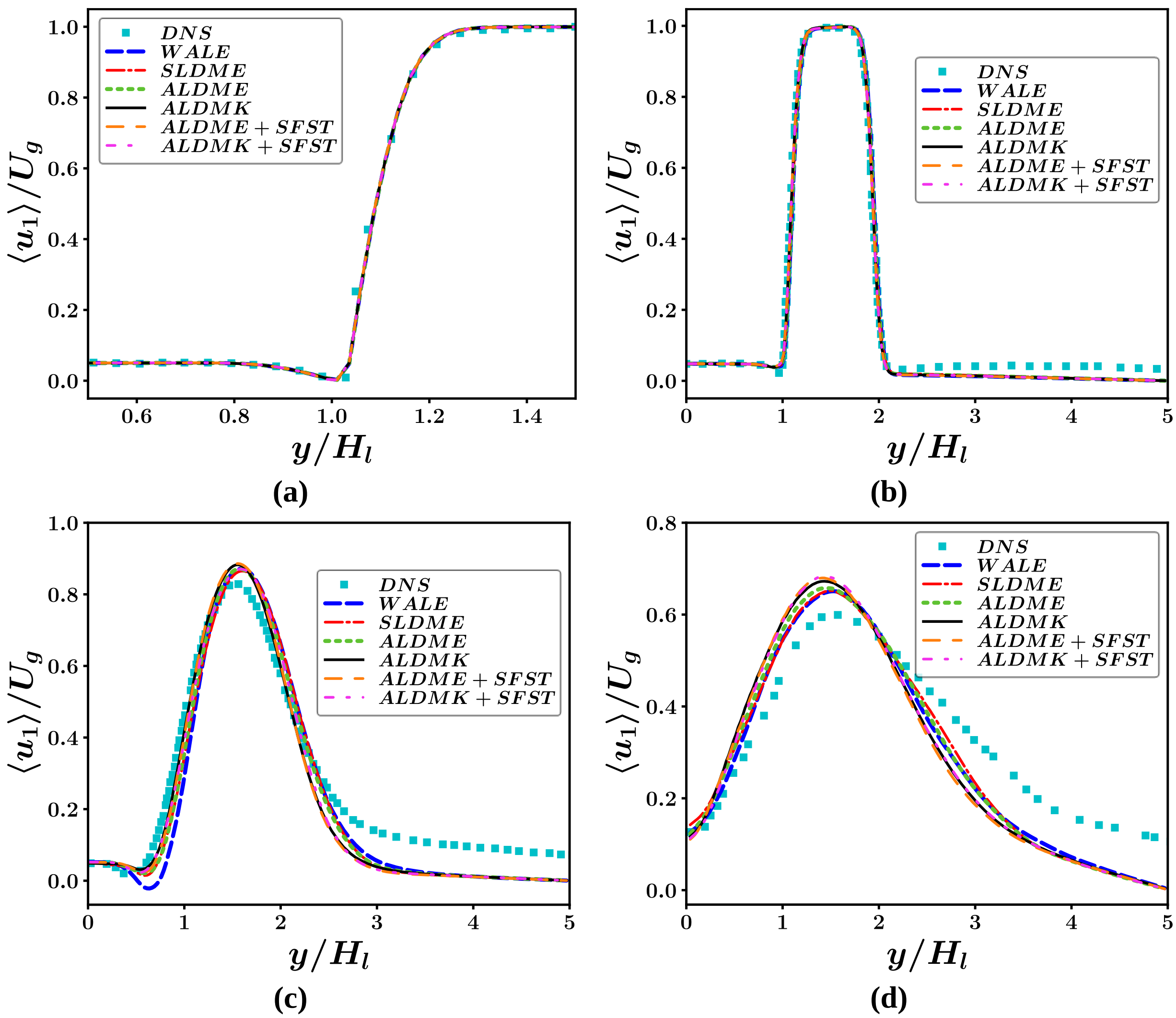

**Figure 8:** The "turbulent two-phase mixing layer" benchmark: comparisons of the mean streamwise velocity at $x = 0.5H_l$ (a), $x = 2H_l$ (b), $x = 6H_l$ (c), and $x = 10H_l$ (d). Different LES closures against the DNS data (Ling *et al.* 2019, Jiang & Ling 2020).

It should be noted that the deviation in the far-field velocity tail (beyond $y/H_l \approx 2.5$) is primarily attributed to the uncertainty of the mathematical formulation of the transverse free boundary condition at the far-field boundary of the domain (at $y = L_y$), rather than the SGS closures themselves. However, all primary atomization dynamics and statistical evaluations are confined to the core mixing region, remaining unaffected by these far-field boundary influence.

To further elucidate the interplay between momentum transfer and interfacial topological changes, the transverse profiles of the mean liquid volume fraction, $\langle \alpha \rangle$, are evaluated in Figure 9.

These profiles track the progressive decay of the liquid core and the spatial distribution of the dispersed phase during primary atomization. The topological evolution observed here is intricately coupled with the shear layer development discussed in the preceding velocity analysis. In the near-field region at $x = 2H_l$ (Figure 9a), the liquid core remains fully intact ($\langle\alpha\rangle \approx 1.0$ for $y < 1.0H_l$). The interface is characterized by a sharp, step-like gradient. Consistent with the mean velocity profiles at this identical location, the interfacial instabilities have not yet transitioned into macroscopic deformations. Consequently, all LES frameworks seamlessly collapse onto the DNS data, accurately preserving the sharp phase boundary without non-physical numerical smearing.

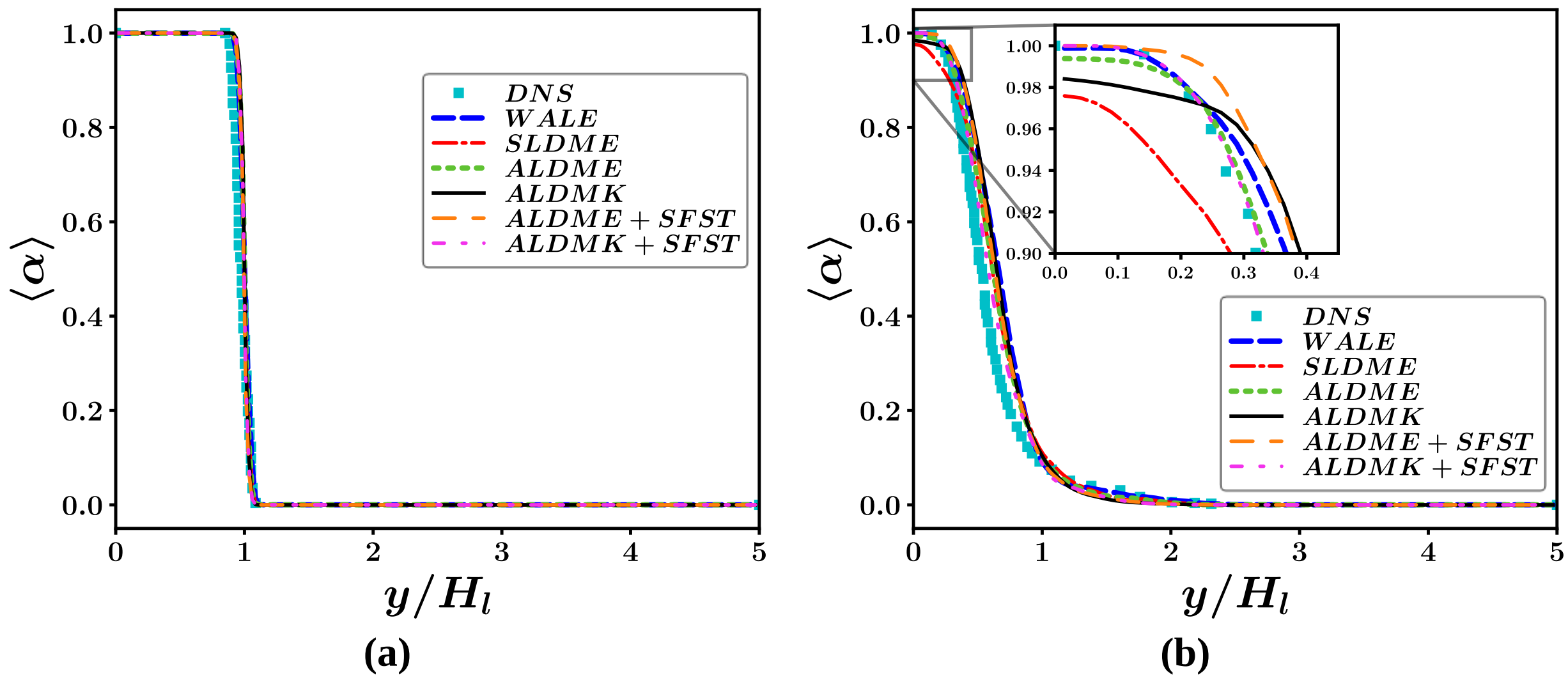

**Figure 9:** The "turbulent two-phase mixing layer" benchmark: comparisons of the mean liquid fraction at $x = 2H_l$ (a) and $x = 6H_l$ (b). Different LES closures against the DNS data (Ling *et al.* 2019).

As the flow advances to $x = 6H_l$ (Figure 9b), the onset of primary atomization broadens the mean liquid fraction profile, reflecting the generation of ligaments and the initial shedding of droplets into the gas stream. At this intermediate stage, the LES models begin to exhibit minor deviations from the DNS benchmark, generally showing a slight delay in the decay of the liquid core (predicting marginally higher liquid fractions in the region $0.2 < y/H_l < 0.8$). However, a distinct and localized discrepancy emerges for the standard dynamic model (S-LDME) in the deep core region ($0 < y/H_l < 0.5$). In this specific zone, the S-LDME framework predicts a reduction in the liquid volume fraction, dropping below unity earlier than both the DNS data and the consistent AD-LES frameworks. This early-stage underprediction reflects the behavior of the standard dynamic procedure in maintaining the structural integrity of the dense phase prior to fully developed atomization.

Furthermore, an evaluation of the liquid fraction profiles at both $x = 2H_l$ (Figure 9a) and $x = 6H_l$ (Figure 9b) reveals that the integration of the SFST closure has a negligible influence during these early to intermediate developmental stages. The profiles produced by the A-LDME+SFST and A-LDMK+SFST models are virtually indistinguishable from their respective baselines. The minute discrepancies observed upon activating the SFST model at these upstream locations are statistically insignificant and can be fundamentally attributed to cumulative numerical/truncation errors rather than distinct physical capillary mechanisms. This confirms that the macroscopic liquid distribution, prior to the far-field chaotic spray formation, is robustly governed by the primary turbulence closures without strict reliance on sub-filter surface tension corrections.

Figure 10 and Figure 11 illustrate the spatial contours of the mean shear and spanwise Reynolds stresses, predicted by the present LES closures versus DNS, providing a direct visual assessment of the transverse turbulent momentum fluxes across the domain. An examination of these contours reveals that all LES closures provide consistent results in terms of the spatial distribution and peak values of the shear Reynolds stress (Figure 10), compared to the DNS. However, the standalone WALE model predicts a lower intensity and spatial spread of the turbulent fluctuations in the spanwise direction (Figure 11). As highlighted by the red bounding box ($0.5 \leq y/H_l \leq 2.5$) specifically delineating the gas-liquid shear layer, the WALE framework underestimates the localized turbulent momentum fluxes compared to the reference DNS data. This underprediction is linked to the numerical dissipation inherent in purely functional EV models, a trend that perfectly elucidates the discrepancies noted in the preceding mean flow analyses. The anomalously high SGS viscosity in the WALE framework dampens the resolved turbulent kinetic energy while simultaneously acting as a spurious scalar diffusivity. Consequently, WALE dampens the energetic eddies required to drive momentum mixing, leading to the flattening of the mean velocity peaks, and diffuses the mean liquid fraction field.

A similar, albeit less severe, deficiency is observed in the standard dynamic formulation (S-LDME). An inspection of the highlighted region (red box) in Figure 11c reveals that the S-LDME model continues to underpredict the intensity of the spanwise Reynolds stress within the primary gas-liquid shear layer. This underestimation is a macroscopic consequence of the structural differences within the standard dynamic procedure. As demonstrated in the channel flow benchmark, the S-LDME framework diminishes the modeled SGS stress contribution in the presence of deconvolution, leading to an insufficient representation of the sub-grid turbulence that is critical for driving interfacial fluctuations.

Conversely, the proposed consistent frameworks—A-LDME and, most notably, the one-equation A-LDMK model—exhibit a substantial improvement in predicting both the spatial distribution and the magnitude of these transverse fluctuations (Figure 11d and e). By maintaining a physically consistent balance and avoiding the non-physical suppression of the modeled stresses, these frameworks more accurately capture the elevated turbulence intensity generated by the intense shear at the gas-liquid interface.

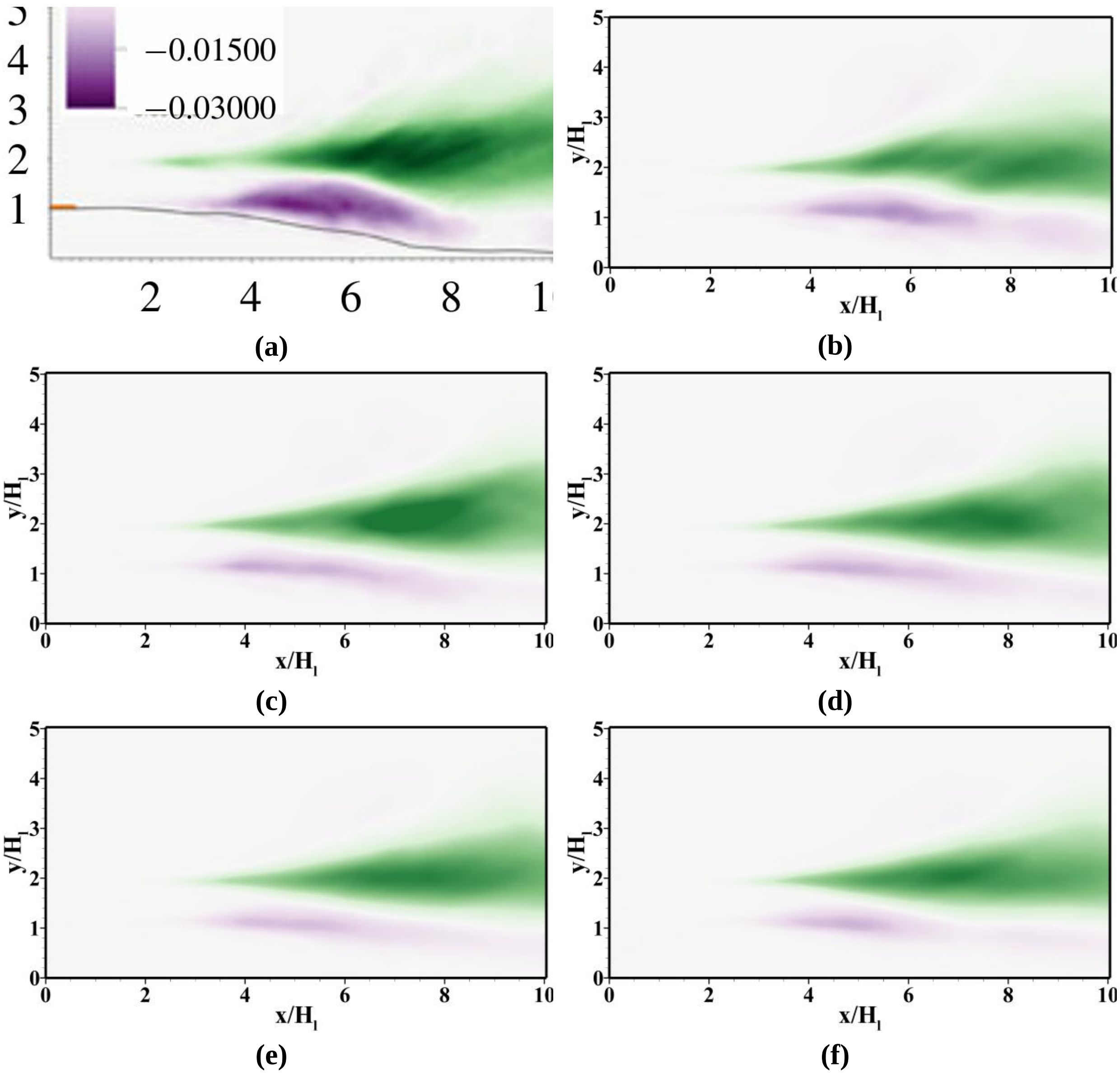

**Figure 10:** The "turbulent two-phase mixing layer" benchmark: comparisons of the total shear Reynolds stress. The DNS data (Ling *et al.* 2019) (a), WALE (b), SLDME (c), ALDME (d), ALDMK (e), and ALDMK+ SFST (f).

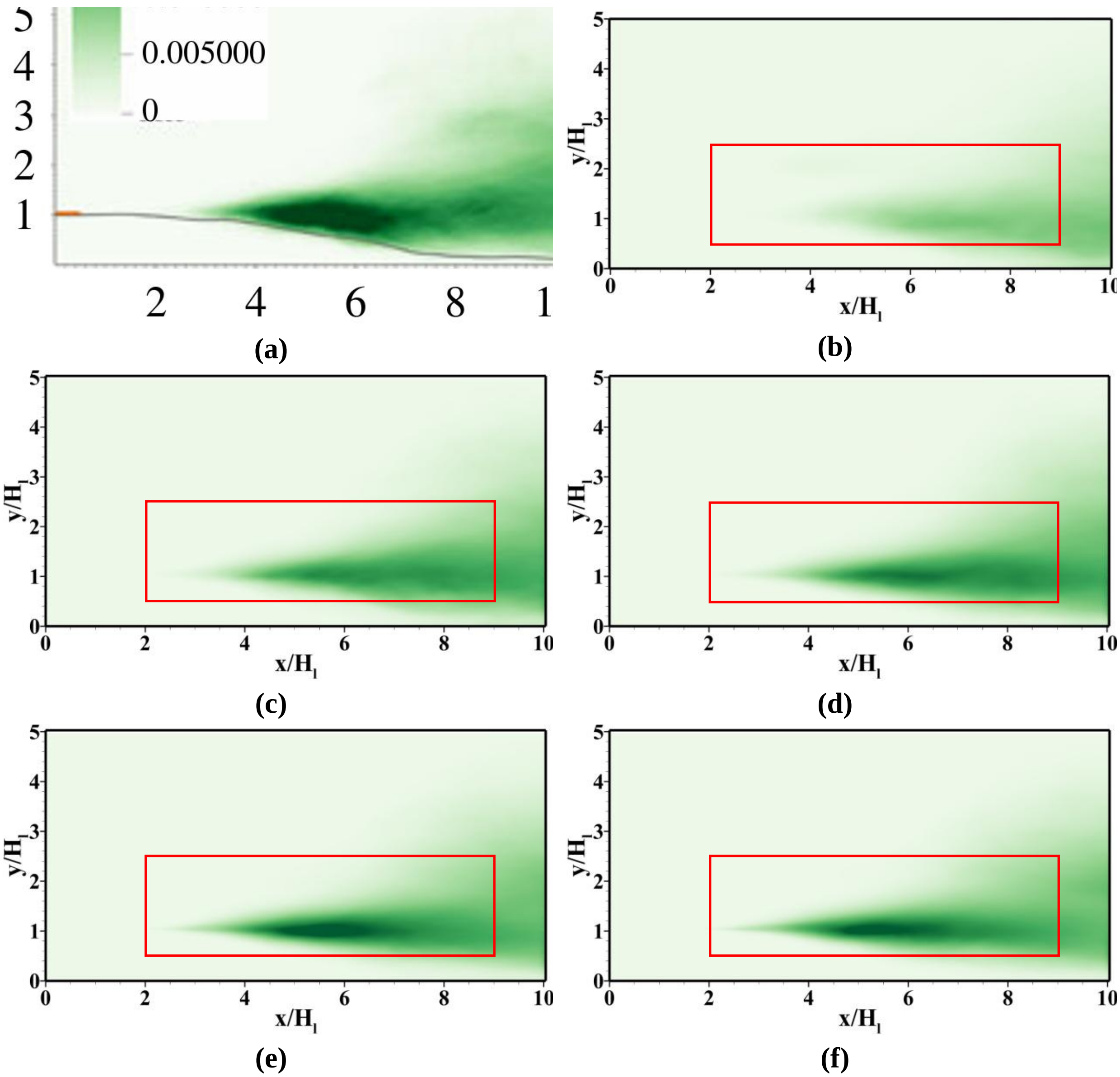

**Figure 11:** The "turbulent two-phase mixing layer" benchmark: comparisons of the total spanwise Reynolds stress. The DNS data (Ling *et al.* 2019) (a), WALE (b), SLDME (c), ALDME (d), ALDMK (e), and ALDMK+ SFST (f).

To further understand the underlying mechanism and confirm the observations from the single-phase channel flow benchmark, the contributions of the deconvolved ($b_{ij}$) and modeled ($a_{ij}$) SFS stress components for the two-phase mixing layer are evaluated. Figure 12 presents the transverse profiles of the shear ($i = 1, j = 2$) and wall-normal ($i = 2, j = 2$) components of these stresses at the downstream location $x = 8H_l$. Consistent with the single-phase results, the standard S-LDME diminishes the modeled SFS stress components ($a_{12}$ and $a_{22}$), reducing them to near zero (red dashed lines in Figure 12b and d). Because the modeled stress is artificially suppressed, the S-LDME model fails to provide sufficient turbulent dissipation through the SGS part, $a_{ij}$.

Consequently, the deconvolved part ($b_{ij}$) attempts to partly compensate for this physical deficit by altering its magnitude (Figure 12a and c). In contrast, the proposed consistent frameworks, A-LDME and A-LDMK, produce more physically meaningful modeled stress contributions and the trace of sub-grid turbulence, which affects the primary atomization cascade in complex multiphase environments.

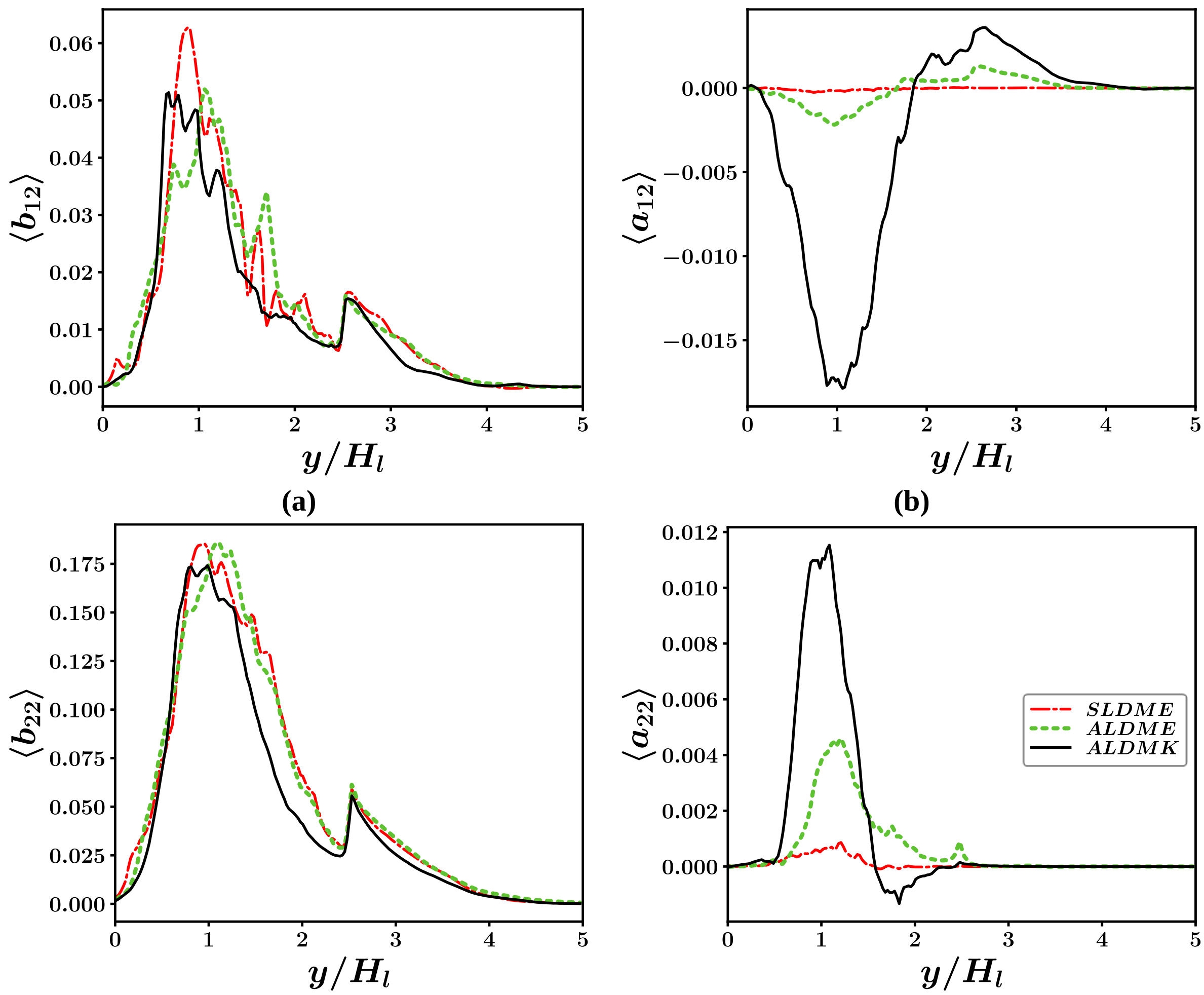

**Figure 12:** The "Mixing layer" benchmark: The comparison of the deconvolved, $b_{ij}$, and modeled, $a_{ij}$, SFS stress components profiles at $x = 8H_l$ using different mixed AD-LES approaches.

To evaluate the scale-resolving capabilities and the dissipative characteristics of the tested SGS closures, the energy spectra of the velocity fluctuations is analyzed. Figure 13 presents the one-dimensional energy spectra for the resolved streamwise ($E_{uu}$) velocity fluctuations. The spectra are extracted along the primary gas-liquid shear layer at a transverse location of $y/H_l = 1$ and two longitudinal locations of ($x/H_l = 4$) and $x/H_l = 6$ to capture the spatial evolution of the turbulence cascade during atomization. The calculated spectra specifically represent the resolved turbulent kinetic energy of the flow. The spectra are computed by applying a one-dimensional Fast Fourier Transform (FFT) along the spanwise (z) direction. To ensure statistical convergence, these instantaneous 1D spectra are ensemble-averaged over time, utilizing all available temporal snapshots from the simulation's sampling period. Finally, the resulting time-averaged spectra are non-dimensionalized: the wavenumbers are scaled by the liquid thickness as $kH_l$, and the energy densities are normalized by $U_g^2 H_l$.

A comparative analysis reveals the differences in the standalone functional approach. The WALE model (blue dashed lines) exhibits an earlier energy drop-off over all flow scales, $kH_l$, especially at intermediate cross sections (Figure 13b). This behavior provides compelling spectral evidence for the excessive numerical dissipation discussed in the preceding momentum and Reynolds stress analyses. By artificially damping the turbulent kinetic energy, mostly at intermediate and small scales, the WALE framework prevents the natural cascade of energy required for realistic interfacial breakdown. In stark contrast, the mixed AD-LES frameworks (S-LDME, A-LDME, and A-LDMK) retain significantly higher energy levels across the resolved wavenumber range.

Furthermore, an examination of the spectral analysis reveals that the isolated impact of the sub-filter surface tension (SFST) closure is relatively marginal. The spectra of the models augmented with the SFST term (orange and magenta lines) remain virtually indistinguishable from their baseline counterparts. While the theoretical intent of explicitly reconstructing sub-grid capillary forces is to resist normal interfacial deformations and penalize premature rupture at the cut-off scale, the current spectral data indicates that this structural modification introduces only minute, inconsequential deviations, at least for this benchmark. Consequently, it is evident that the energy cascade and the ultimate dissipation of smaller coherent structures are predominantly governed by the primary turbulence momentum closures, with the SFST exerting no substantial influence on the overall spectral energy distribution at this resolution.

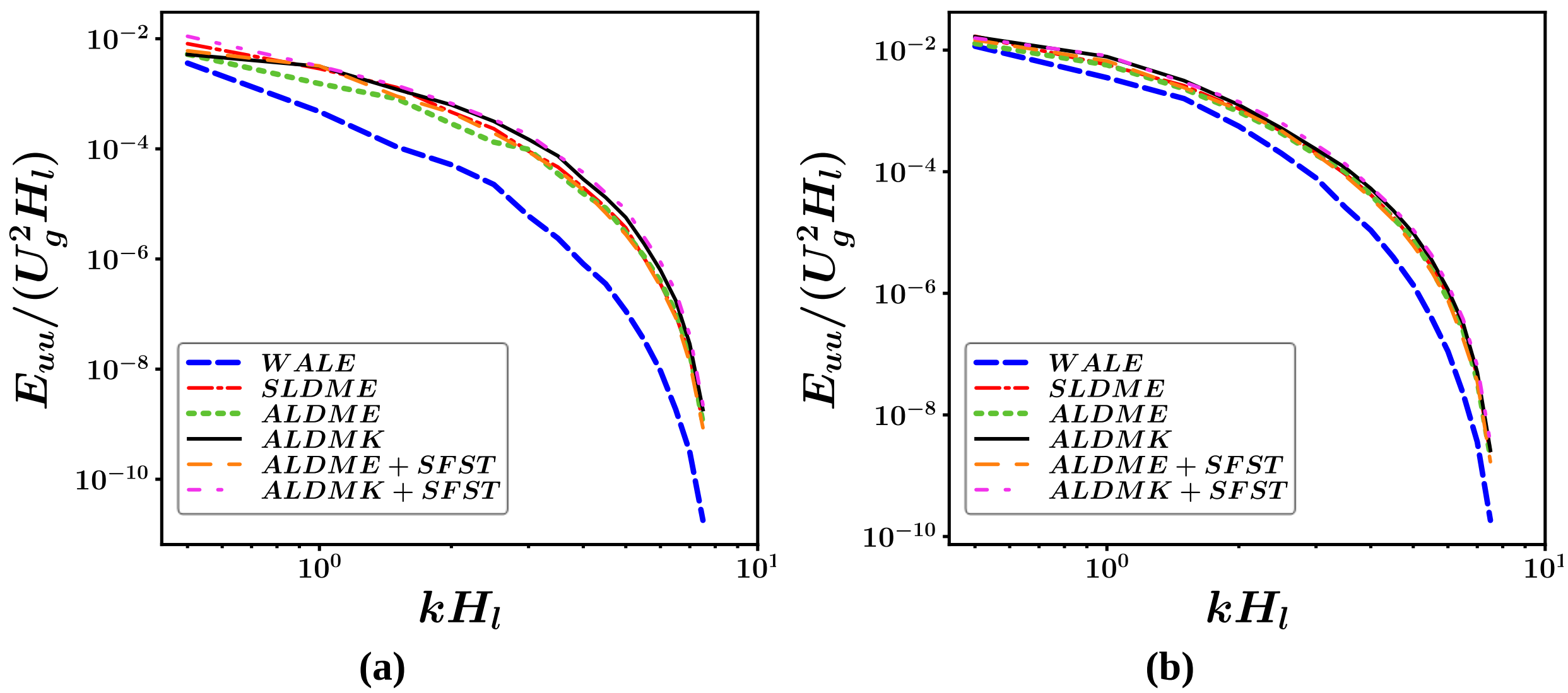

**Figure 13:** The "turbulent two-phase mixing layer" benchmark: The 1D energy spectra of the resolved streamwise velocity fluctuations at $y/H_l = 1$, and various streamwise locations: $x/H_l = 4$ (a), and $x/H_l = 6$ (b). Comparisons of different LES predictions.

To quantify the atomization characteristics and evaluate the model's capability in capturing the fragmentation process, the Droplet Size Distribution (DSD) is analyzed in the downstream region of the mixing layer. Following the methodology established in the reference DNS study by Ling et al. (2017), the sampling is performed within a cubic control volume located in the developed spray region ($8 \leq x/H_l \leq 10$, $0.5 \leq y/H_l \leq 2.5$), spanning the entire spanwise width of the domain.

The droplet statistics are presented in terms of the average number of droplets, $n_d(d)$, normalized by the total number of samples, $N_s$, i.e., $n_d(d) = N(d)/N_s$ where $N(d)$ represents the total count of droplets collected within a bin centered at each nominal diameter $d$. To ensure statistical reliability, particularly for larger droplets which are statistically scarce, a variable bin size strategy is employed, starting from a minimum width of 6.25 μm and increasing with a constant geometric ratio of 1.2. Furthermore, due to the limitations of the VOF method, it is important to distinguish between resolved and under-resolved structures, as discussed by Ling et al. (2017). Droplets with diameters smaller than twice the grid spacing ($d < 2\Delta$) are often prone to numerical errors and artificial coalescence/breakup. Therefore, while the full spectrum is presented, the analysis primarily focuses on the statistically trustworthy range of $d \geq 2\Delta$.

Figure 14 presents the histograms of the normalized droplet number density, $n_d(d)$, obtained from the different mixed AD-LES closures compared against the reference DNS data. The time-averaging period for each simulation (>54 ms) is long enough to achieve a low level of statistical error, with statistical uncertainties (for all models) below 0.5% and 3.5% for the smallest (13.75 ≤ $d \leq 22.75\ \mu m$) and largest ($247.4 \leq d \leq 303.1\ \mu m$) droplet bins, respectively. As established in the methodology, the analysis is strictly confined to the fully-resolved trustworthy range of $d \geq 2\Delta$ (i.e., $d \geq 50\ \mu m$) to exclude under-resolved scales prone to numerical artifacts.

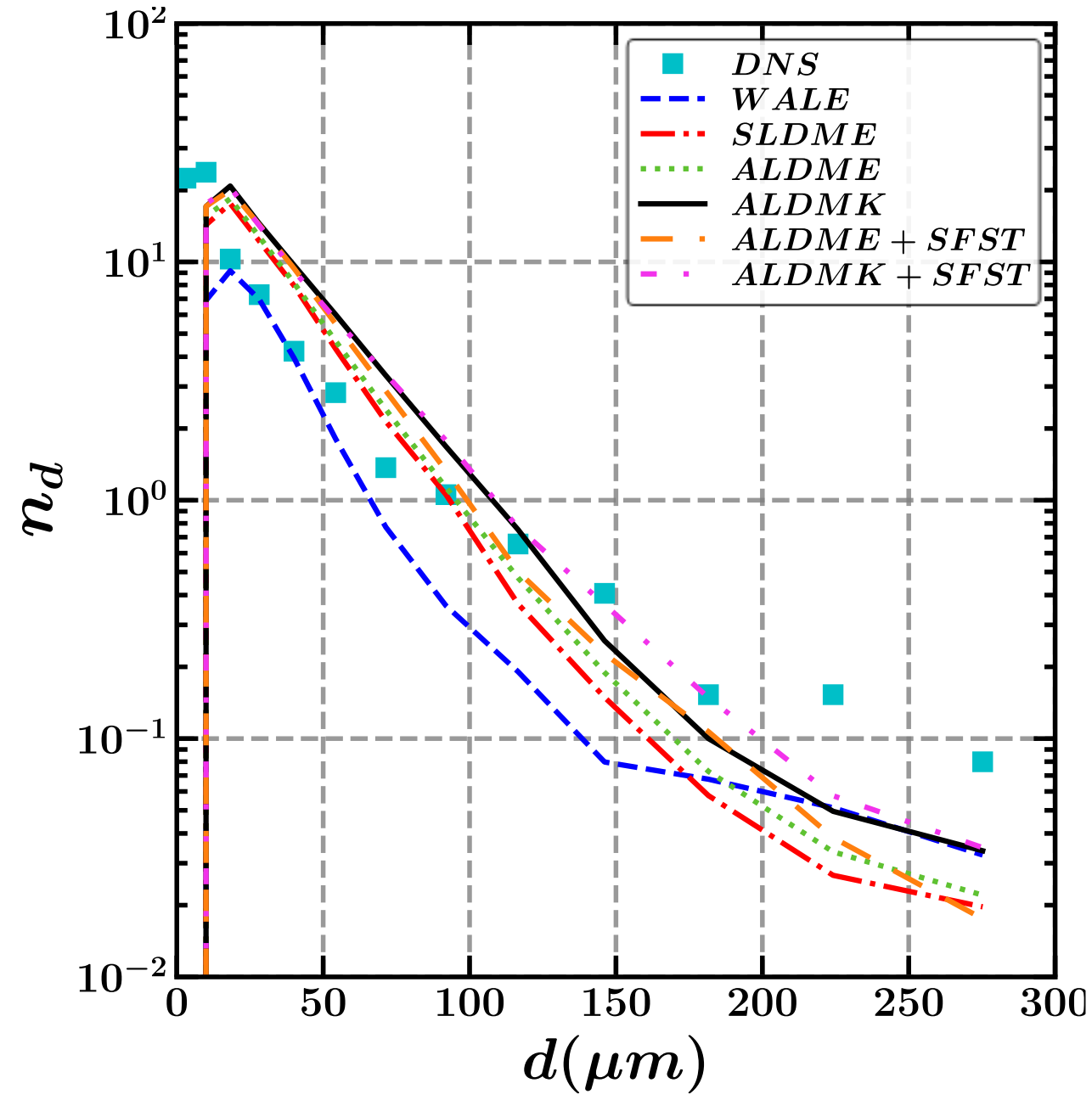

**Figure 14:** The "turbulent two-phase mixing layer" benchmark: comparisons of droplet size distribution predicted by different LES closures against the DNS data (Ling *et al.* 2017).

While the standalone WALE model appears to yield closer agreement for droplets smaller than 70 $\mu$m, its accuracy decreases for sizes larger than 100 $\mu$m. In addition, generally, the predictions of the smallest droplet-size groups are less reliable due to numerical uncertainties arising for unresolved droplets, as the droplet sizes become comparable to the grid resolution. The anomalously high SGS viscosity suppresses the resolved turbulent kinetic energy, effectively dampening the critical shear-driven interfacial instabilities required to initiate primary atomization. Consequently, instead of resolving the physical fragmentation cascade, the WALE framework non-physically diffuses the gas-liquid interface, preventing the natural formation of ligaments and resulting in a severely truncated DSD that fails to capture the continuous spectrum of droplet sizes.

Similarly, the DSD predicted by the standard mixed model, S-LDME, exhibits a significant deviation from the DNS benchmark, especially at larger droplet sizes. While the model captures the presence of smaller droplets near the cut-off scale, it displays a precipitous decay in the population of intermediate and large droplets ($d > 100$ μm). The histogram reveals a marked under-prediction of droplet counts in the $100 - 200$ μm range and a near-total absence of droplets larger than 200 μm. This behavior is intrinsically coupled to the model's limitations observed in the preceding analyses. As demonstrated in the Reynolds stress contours (e.g., Figure 11c), the S-LDME formulation diminishes the modeled SGS stresses, generating lower turbulent momentum fluxes to drive the ligament breakup process at the interface. Because it cannot sustain the appropriate interfacial shear, the liquid core undergoes the artificial early decay previously observed in the mean scalar profiles (Figure 9b, $y/H_l < 0.5$). Consequently, instead of experiencing a physical cascade where the continuous liquid phase breaks into large ligaments and subsequently into a broad spectrum of intermediate and small droplets, the liquid core tends to fragment artificially into smaller, grid-scale structures, entirely missing the broad size spectrum characteristic of fully developed primary atomization.

The transition to the consistent A-DEV framework via the algebraic A-LDME model yields a slight improvement in the droplet size distribution. This enhancement is intrinsically tied to the recovery of the resolved turbulent momentum fluxes documented in the earlier stress analysis. As observed in the spanwise Reynolds stress contours (e.g., Figure 11d), the consistent reconstruction of the SGS stress tensor prevents the artificial suppression of interfacial turbulence. Consequently, this sustained physical shear successfully mitigates the premature numerical decay of the liquid core (as previously shown in Figure 9b), promoting more realistic macroscopic ligament formation rather than instantaneous diffusion. As a result, the A-LDME histogram displays a wider spread, with a recoverable population of intermediate droplets. However, a quantitative deficit persists; the probability density of larger droplets remains lower than the DNS reference. This limitation implies that the local equilibrium assumption inherent in the algebraic closure—which instantaneously balances sub-grid production and dissipation—may still be overly dissipative in the highly transient, non-equilibrium regions where primary breakup occurs.

The closest agreement with the DNS data among the tested models is achieved by the A-LDMK model. In the statistically reliable diameter range, the A-LDMK distribution closely mirrors the characteristic profile of the DNS, capturing not only the decay rate of smaller droplets but also the crucial "tail" of the distribution corresponding to larger droplets, in a closer agreement to the DNS. This improved performance supports the inclusion of the SGS turbulent kinetic energy transport equation. Unlike the algebraic closures, A-LDMK explicitly accounts for flow history and non-equilibrium effects. Although the subgrid TKE equation in the present framework is kinematically adapted and its production is driven by shear, it allows the spatial transport and temporal memory of this shear-generated TKE from production sites to regions of intense interfacial interaction. This transported kinetic energy sustains the physical instabilities required for primary atomization and reproduces the complex energy backscatter, sustaining the mean velocity peaks far downstream (Figure 8). This physical fidelity enables the A-LDMK framework to maintain the high-intensity turbulent fluctuations (Figure 11e) strictly required to physically disintegrate macroscopic liquid ligaments, culminating in a droplet size distribution that is statistically comparable to the high-fidelity DNS data.

The ultimate macroscopic manifestation of modeling sub-filter capillarity is observed in the droplet size statistics. As illustrated in Figure 14, augmenting the consistent frameworks with the SFST closure (A-LDME+SFST and A-LDMK+SFST) introduces a level of improvement in the droplet count probabilities, especially over the intermediate range of droplet size span, compared to their baseline counterparts. The level of improvement is more significant compared to the previous velocity statistics. This indicates that the reconstructed sub-grid surface tension term, $b_i^{\mu\sigma}$, exerts stronger influence during the intermediate stages of the fragmentation cascade at the current computational resolution. To conclude this analysis, the primary atomization dynamics and the resulting droplet size spectrum are predominantly governed by the core SGS momentum closures, and the consistent A-LDMK work better than the other closures in predicting this phenomenon. However, the inclusion of SFST effect, which is theoretically formulated to provide structural resistance against SGS topological deformations, can also improve the DSD predictions.

Finally, for a quantitative assessment of the performance of the evaluated closures in predicting the primary flow statistics for the two-phase flow benchmark, the $L_2$ error norms for the mean streamwise velocity and mean liquid volume fraction are computed with respect to the DNS data, over all profiles at $x/H_l$ =2, 4, 6, and 8, within the region ($2 \leq x/H_l \leq 8$, $0.5 \leq y/H_l \leq 2.5$) shown by the red box in Figure 11. For $\langle \alpha \rangle$, the section $x/H_l = 2$ is excluded since all results closely match each other (see Figure 9a). The rationale for selecting this region is threefold. First, this region is the upstream zone of the spray region ($8 \leq x/H_l \leq 10$, $0.5 \leq y/H_l \leq 2.5$) used here and in the DNS for computing droplet statistics (discussed earlier). Second, it contains important portions of the liquid-gas interface (see Figure 1). Third, this region is far from the far-field boundary of the domain to avoid the uncertainties imposed by this boundary. The calculated error norms, relative to the reference DNS data, are summarized in Table 6.

**Table 6:** The "turbulent two-phase mixing layer" benchmark: The $L_2$ error norms of the predicted mean streamwise velocity and mean liquid volume fraction, within region shown by the red box in Figure 11.

| **Statistics** | **WALE** | **SLDME** | **ALDME** | **ALDMK** | **ALDME +SFST** | **ALDMK +SFST** |
|---|---|---|---|---|---|---|
| $\langle u_1 \rangle$ | 0.11 | 0.08 | 0.07 | 0.07 | 0.08 | 0.07 |
| $\langle \alpha \rangle$ | 0.34 | 0.28 | 0.31 | 0.40 | 0.41 | 0.33 |

For the mean streamwise velocity profiles, $\langle u_1 \rangle$, the average prediction errors by all new AD-LES models are about 7% compared to 8% for S-LDME and 11% for the WALE model. For the mean volume fraction, $\langle \alpha \rangle$, S-LDME and A-LDME have the smallest average errors while all errors lie between about 30 to 40%.

## 6. Conclusion

In this study, a mathematically consistent mixed AD framework was extended to turbulent two-phase flows to address the fundamental computational trade-off between numerical robustness and physical fidelity in predicting primary atomization. Initially, the standard mixed EV-AD (S-LDME) and baseline purely functional EV WALE models were evaluated, revealing critical deficiencies in their SGS representations. To resolve these inconsistencies, a one-equation closure, namely A-LDMK, was developed to incorporate non-equilibrium flow history. This was coupled with a structural SFST formulation designed to reconstruct unresolved capillary forces. The predictive capabilities and numerical stability of these closures were assessed *a posteriori* through a single-phase turbulent channel flow and a complex turbulent gas-liquid mixing layer benchmark.

The results demonstrated that the standalone functional model (WALE) exhibited higher numerical dissipation. This over-diffusion artificially suppresses critical shear-driven interfacial instabilities, non-physically smears the phase boundaries, and ultimately fails to reproduce accurate distribution of droplet sizes. Furthermore, an inconsistency in the conventional S-LDME closure was shown *a posteriori* in both single- and two-phase flow benchmarks. By unphysical vanishing of the modeled sub-grid stress contribution, S-LDME precipitated a premature numerical decay of the liquid core. From a numerical stability standpoint, accommodating physical energy backscatter in the highly transient multiphase regime proved uniquely challenging for S-LDME, necessitating a severely restrictive *ad hoc* clipping threshold ($\nu_r/\nu \geq -0.5$) that artificially truncated essential flow dynamics.

Conversely, the A-LDME and A-LDMK AD-LES frameworks addressed these issues. The algebraic formulation (A-LDME) demonstrated exceptional inherent robustness; in the single-phase channel flow benchmark, it remained entirely stable without requiring any *ad hoc* clipping constraints, while accommodating for large backscatters proportional to $\nu_r/\nu \approx -4$. Moreover, it offered higher computational efficiency by circumventing the need to solve an additional transport equation for the SGS kinetic energy. In the complex multiphase regime, both A-LDME and A-LDMK required only the standard Positive Total Viscosity (PTV) limit, proving far superior to the standard S-LDME in preserving local energy backscatter. Compared to the primary role of the SGS momentum, the inclusion of SFST in the consistent closures had secondary effects on the prediction of the velocity statistics, while a level of improvement was observed in the predicted droplet size distribution, particularly in the intermediate stage of atomization. It should be noted that the scope of the current work was limited to the subgrid momentum and surface tension closures. The structural reconstruction of the subgrid phase transport term was found to induce numerical instability. Modifying the numerical framework to safely incorporate this subfilter phase flux remains an important subject for future investigations.

Ultimately, by accounting for the non-equilibrium kinetic energy transfer and preserving the high-intensity interfacial shear, the one-equation A-LDMK framework provided the closest agreement with the reference data in capturing the primary atomization cascade and the droplet size distribution among the evaluated models.

## Nomenclature

### Mathematical Operators & Accents

- $\partial_t$: Time derivative ($\partial/\partial t$)
- $\partial_i$: Spatial derivative ($\partial/\partial x_i$)
- $\widetilde{(.)}$: Implicit grid filter
- $\overline{(.)}$: Explicit filtering operation
- $G * (.)$: Explicit filtering operation
- $G^{-1}$: Inverse explicit filter operator
- $\widehat{(.)}$: Explicit test filtering
- $\breve{(.)}$: Test box filter
- $(.)^*$: (Deconvolved) approximation
- $\langle . \rangle$: Ensemble (Reynolds) averaging
- $(.)'$: Fluctuating component
- $|.|$: Magnitude or norm of a tensor/vector

### Roman Symbols

- $a$: Modeled SFS component
- $b$: Deconvolved SFS component
- $C_D$: DEV coefficient/dynamic coefficient

- $C_e$: Model constant
- $g_i$: Gravity vector
- $k$: Turbulent kinetic energy
- $k^*$: Model-level TKE
- $L_{ij}$: Leonard stress
- $m_{ij}$: Tensor to which the Germano-Lilly identity is applied.
- $M_{ij}$: Tensor defined based on the chosen DEV closure
- $n_i$: Interface normal vector
- $N$: Number of deconvolution iterations
- $p$: Mixture pressure.
- $P_{ij}$: Tensor used in the dynamic coefficient calculation
- $q^*$: Turbulent kinetic energy at the model level
- $\widehat{q^*}$: Turbulent kinetic energy at the test-filter level
- $Q_N$: Numerical approximation of $G^{-1}$
- $S$: Characteristic strain rate magnitude
- $S_{ij}$: Strain-rate tensor
- $t$: Time
- $u_i$: (Mixture) velocity vector
- $\langle u_i' u_j' \rangle$: Reynolds stress
- $x_i$: Spatial coordinate

**Greek Symbols**

- $\alpha$: Primary (liquid) phase volume fraction
- $\gamma$: Exponent used in the SFS viscosity formulation
- $\delta_{ij}$: Kronecker delta
- $\Delta$: Local LES grid size, filter width
- $\kappa$: Interface curvature
- $\mu$: Molecular viscosity
- $\mu_r$: SFS dynamic viscosity
- $\nu_r$: SFS kinematic viscosity
- $\rho$: Mixture density
- $\sigma$: Surface tension coefficient
- $\sigma_k$: turbulent Prandtl/Schmidt number
- $\tau_{ij}$: SGS correlation tensor
- $\tau_i$: SGS correlation vector
- $\phi$: A generic variable
- $\omega_C'$: Dimensionless cut-off wavenumber
- $\omega_N'$: Dimensionless Nyquist wavenumber

**Subscripts & Superscripts**

- $l$: Primary (liquid) phase properties (e.g., $\rho_l, \mu_l$)
- $v$: Vapor phase properties (e.g., $\rho_v, \mu_v$)
- $r$: Deviatoric component of a tensor (e.g., $a_{ij}^{uu,r}$)
- $T$: Test-level formulation (e.g., $\tau_{ij}^{T,uu}$)
- $i, j, n$: Spatial directions/Tensor and vector indices (e.g., $x_i, u_j$)
- $uu$: Momentum inertia correlation term (e.g., $T_{ij}^{uu}, a_{ij}^{uu}$)
- $u\sigma$: Surface tension correlation term (e.g., $T_i^{u\sigma}, a_i^{u\sigma}$)
- $\alpha u$: Phase transport correlation term (e.g., $T_i^{\alpha u}, a_i^{\alpha u}$)
- $u\mu$: Viscous stress correlation term (e.g., $\tau_{ij}^{u\mu}$)
- $u\rho$: Density-velocity correlation term (e.g., $\tau_i^{u\rho}$)
- $SGS$: Sub-Grid Scale

- $SFS$: Sub-Filter Scale

## Appendix A: Derivation of the sub-filter turbulent kinetic energy transport equation

Similar to the approach followed in the pioneering wroks by (Lilly 1967) or (Deardorff 1973) for a single-phase incompressible flow, the exact transport equation for the total kinetic energy, $K = \frac{1}{2}u_i u_i$, is derived by taking the scalar dot product of the momentum equation of the one-fluid formulation for two-phase flows, Eq. (1), with the velocity vector $u_i$. Then, applying the spatial grid filter operator yields the filtered total kinetic energy equation:

$$\partial_t(\widetilde{\rho K}) + \partial_j\left(\widetilde{\rho u_j K}\right) = -\partial_i(\widetilde{p u_i}) + \partial_j\left(2\widetilde{\mu u_i S_{ij}}\right) - \left(2\widetilde{\mu S_{ij} S_{ij}}\right) + (\widetilde{\sigma\kappa u_i \partial_i \alpha}) + (\widetilde{\rho u_i}) g_i, \qquad \text{(A1)}$$

Similarly, the resolved kinetic energy, defined as $K^R = \frac{1}{2}\tilde{u}_i\tilde{u}_i$, is governed by an equation obtained by taking the scalar dot product of the filtered momentum equation (Eq. (8)) with the resolved velocity $\tilde{u}_i$. This operation requires the resolved-scale mass continuity equation, $\partial_t\tilde{\rho} + \partial_i(\tilde{\rho}\tilde{u}_i) + \partial_i\tau_i^{u\rho} = 0$, where $\tau_i^{u\rho}$ is the sub-grid density-velocity correlation. The sub-grid kinetic energy ($k_{\text{SGS}}$) is defined by the difference between the filtered total kinetic energy and the resolved kinetic energy, such that $\tilde{\rho}k_{\text{SGS}} = (\widetilde{\rho K}) - \tilde{\rho}K^R$. Subtracting the resolved kinetic energy ($K^R$) equation from the filtered total kinetic energy equation, (A1), yields the exact transport equation for $k_{\text{SGS}}$:

$$\begin{aligned}
\partial_t(\tilde{\rho}k_{\text{SGS}}) + \partial_j\left(\tilde{\rho}\tilde{u}_j k_{\text{SGS}}\right) = &-\partial_j \underbrace{\left[\left(\widetilde{\rho u_j k}\right) - \tilde{u}_j(\widetilde{\rho k})\right]}_{\text{Turbulent Diffusion}} - \partial_i \underbrace{\left[(\widetilde{p u_i}) - \tilde{p}\tilde{u}_i\right]}_{\text{Pressure Diffusion}} + \\
&\partial_j \underbrace{\left[\left(2\widetilde{\mu u_i S_{ij}}\right) - 2\tilde{\mu}\tilde{S}_{ij}\tilde{u}_i + \tau_{ij}^{uu}\tilde{u}_i - \tau_{ij}^{u\mu}\tilde{u}_i\right]}_{\text{Viscous \& SGS Stress Diffusion}} - \underbrace{\left[\left(2\widetilde{\mu S_{ij} S_{ij}}\right) - 2\tilde{\mu}\tilde{S}_{ij}\tilde{S}_{ij}\right]}_{\varepsilon_{\text{SGS}}\ \text{(Dissipation)}} - \\
&\underbrace{\tau_{ij}^{uu}\tilde{S}_{ij}}_{P_{\text{SGS}}\ \text{(Production)}} + \underbrace{\tau_{ij}^{u\mu}\tilde{S}_{ij}}_{\text{Viscosity Fluctuation Work}} + \underbrace{\left[(\widetilde{\sigma\kappa u_i \partial_i \alpha}) - \sigma\tilde{\kappa}\tilde{u}_i\partial_i\tilde{\alpha} - \tilde{u}_i\tau_i^{u\sigma}\right]}_{\text{SGS Surface Tension Work}} + \\
&\underbrace{\tau_i^{u\rho} g_i}_{\text{Buoyancy Production}} - K^R\partial_j\tau_j^{u\rho} + \tilde{u}_i\partial_t\tau_i^{u\rho},
\end{aligned} \qquad \text{(A2)}$$

To achieve a closed form, several modeling assumptions can be applied:

- The terms representing the density-velocity correlation ($\tau_j^{u\rho}$), viscous stress fluctuations ($\tau_{ij}^{u\mu}$), and SGS surface tension work are assumed negligible in Eq. (A2).
- The SGS stress tensor $\tau_{ij}^{uu}$ is modeled using the Boussinesq EV approximation, $\tau_{ij}^{uu} - \frac{1}{3}\tau_{kk}^{uu}\delta_{ij} = -2\mu_r\tilde{S}_{ij}$. Because the resolved field is divergence-free ($\tilde{S}_{ii} = 0$), the production term simplifies to $P_{\text{SGS}} = 2\mu_r\tilde{S}_{ij}\tilde{S}_{ij} = \mu_r\tilde{S}^2$.

- The viscous dissipation at small scales is modeled based on Kolmogorov's hypothesis and a dimensional analysis as $\varepsilon_{\mathrm{SGS}} = C_\epsilon \frac{\tilde{\rho} k_{\mathrm{SGS}}^{3/2}}{\tilde{\Delta}}$, assuming a modeling constant of $C_\epsilon = 1$.
- The combined turbulent, pressure, and viscous diffusion terms are approximated using the gradient diffusion hypothesis, $D_{\mathrm{SGS}} = \partial_j \left[ \left( \mu + \frac{\mu_r}{\sigma_k} \right) \partial_j k_{\mathrm{SGS}} \right]$, assuming a turbulent Schmidt number of $\sigma_k = 1$.

Applying these closures yields the standard modeled transport equation:

$$\partial_t(\tilde{\rho} k_{\mathrm{SGS}}) + \partial_j(\tilde{\rho} \tilde{u}_j k_{\mathrm{SGS}}) = \partial_j[(\mu + \mu_r)\partial_j k_{\mathrm{SGS}}] + \mu_r S^2 - \frac{\tilde{\rho} k_{\mathrm{SGS}}^{3/2}}{\tilde{\Delta}}, \tag{A3}$$

In the A-LDMK framework, the conventionally filtered variables are replaced by their deconvolved approximations obtained via the Van Cittert algorithm. Therefore, substituting $\tilde{u}_i$ with $u_i^\star$ and $\tilde{S}^2$ with $S^{\star 2} = 2 S_{ij}^\star S_{ij}^\star$ into Eq. (A3) yields:

$$\partial_t(\rho^\star k^\star) + \partial_j(\rho^\star u_j^\star k^\star) = \partial_j[(\mu + \mu_r)\partial_j k^\star] + \mu_r S^{\star 2} - \frac{\rho^\star {k^\star}^{3/2}}{\tilde{\Delta}} \tag{A4}$$

From a numerical implementation perspective within the OpenFOAM finite volume framework, the transport equation for the sub-filter scale turbulent kinetic energy, Eq. (A4) requires careful treatment to ensure solver stability. The transient, convective, and diffusive terms are discretized implicitly to maintain a stable integration over time. Specifically, for the convective term, the deconvolved mass flux ($\rho^\star u_j^\star$), instead of $\tilde{\rho}\tilde{u}_j$, is utilized to transport the $k^*$ field. The production term, which is driven by the resolved scales and depends on the deconvolved strain rate tensor ($S^*$) and the sub-grid EV ($\mu_r$), is updated based on the available fields at each iteration of the momentum equation and treated as an explicit source term.

A critical aspect of the numerical implementation is the treatment of the highly non-linear dissipation term (the sink term). Solving this term purely explicitly often leads to numerical divergence due to rapid localized energy depletion. To prevent numerical instability and maintain the diagonal dominance of the resultant algebraic matrix, this sink term is mathematically linearized and discretized implicitly. In the developed solver, the term $-\rho^*(k^*)^{3/2}/\tilde{\Delta}$ is factored into a linearized coefficient multiplied by the unknown variable $k^*$, implemented via the fvm::Sp operator in OpenFOAM as $-(\rho^*\sqrt{k^*}/\tilde{\Delta})k^*$. This semi-implicit treatment strongly enhances the robustness of the dynamic procedure.

Finally, to satisfy the physical realizability conditions and mathematically guarantee that the SGS kinetic energy remains strictly non-negative throughout the multiscale interactions, a bounding procedure is enforced. Subsequently, after the matrix is solved at each time step, the $k^*$ field is explicitly bounded to a minimum threshold, set to a marginally small positive value, to strictly eliminate any non-physical negative values.

## Data availability

The data that support the findings of this study are available from the corresponding author upon reasonable request.

**Declaration of Interests:** The authors report no conflict of interest

## Acknowledgments

This work was partially supported by TÜBITAK [grant number 221M421].